\documentclass[a4paper,11pt]{article}
\pdfoutput=1 

\usepackage{jheppub} 

\usepackage[T1]{fontenc} 
\usepackage{arydshln}

\usepackage{rotating}

\allowdisplaybreaks

\def\beq{\begin{equation}}
\def\eeq#1{\label{#1}\end{equation}}
\def\eeqn{\end{equation}}

\def\beqa{\begin{eqnarray}}
\def\eeqa#1{\label{#1}\end{eqnarray}}
\def\eeqan{\end{eqnarray}}
\def\CR{\nonumber \\ }

\def\bseq{\begin{subequations}}
\def\eseq#1{\label{#1}\end{subequations}}
\def\eseqn{\end{subequations}}

\def\Dlr{\overleftrightarrow{D}}
\def\VCKM{V_{\text{CKM}}}
\def\L{{\cal L}}
\def\O{{\cal O}}
\def\M{{\cal M}}
\def\IBP{\text{IBP}}
\def\EoM{\text{EoM}}
\def\SM{\text{SM}}
\def\NP{\text{NP}}

\def\BE{B$_\text{E}$}
\def\BS{B$_\text{S}$}

\def\cw{c_\theta}
\def\sw{s_\theta}
\def\cwb{\bar{c}_\theta}
\def\swb{\bar{s}_\theta}
\def\cwh{\hat{c}_\theta}
\def\swh{\hat{s}_\theta}
\def\Dgzb{\Delta\bar g_1^Z}
\def\Dgab{\Delta\bar g_1^\gamma}
\def\Dkapzb{\Delta\bar\kappa_Z}
\def\Dkapab{\Delta\bar\kappa_\gamma}
\def\lamzb{\bar\lambda_Z}
\def\lamab{\bar\lambda_\gamma}
\def\lamgb{\bar\lambda_g}
\def\DkapFb{\Delta\bar\kappa_F}
\def\DkapVb{\Delta\bar\kappa_V}
\def\De{\Delta\epsilon}
\def\dgL{\delta g_L}
\def\dgR{\delta g_R}
\def\dgz{\delta g_{1z}}
\def\dkapa{\delta\kappa_\gamma}
\def\dkapz{\delta\kappa_z}

\def\dbNP{\bar\delta^{\text{NP}}}
\def\Obs{\hat\O}

\preprint{MCTP-15-24}

\title{\boldmath Effective theories of universal theories}

\author{James~D.~Wells}
\author{and Zhengkang~Zhang}
\affiliation{Michigan Center for Theoretical Physics, Department of Physics, University of Michigan,\\Ann Arbor, MI 48109, U.S.A.}

\emailAdd{jwells@umich.edu}
\emailAdd{zzkevin@umich.edu}

\abstract{It is well-known but sometimes overlooked that constraints on the oblique parameters (most notably $S$ and $T$ parameters) are generally speaking only applicable to a special class of new physics scenarios known as universal theories. In the effective field theory (EFT) framework, the oblique parameters should not be associated with Wilson coefficients in a particular operator basis, unless restrictions have been imposed on the EFT so that it describes universal theories. We work out these restrictions, and present a detailed EFT analysis of universal theories. We find that at the dimension-6 level, universal theories are completely characterized by 16 parameters. They are conveniently chosen to be: 5 oblique parameters that agree with the commonly-adopted ones, 4 anomalous triple-gauge couplings, 3 rescaling factors for the $h^3$, $hff$, $hVV$ vertices, 3 parameters for $hVV$ vertices absent in the Standard Model, and 1 four-fermion coupling of order $y_f^2$. All these parameters are defined in an unambiguous and basis-independent way, allowing for consistent constraints on the universal theories parameter space from precision electroweak and Higgs data.}

\begin{document} 
\maketitle
\flushbottom

\section{Introduction}

It has been realized for quite some time now that precision measurements of Standard Model (SM) processes can provide indirect probes of beyond the Standard Model (BSM) new physics. Over the past few decades, high-precision measurements of electroweak and flavor observables have found remarkable agreement with the SM, leading to stringent constraints on BSM effects in these sectors; see e.g.~\cite{Agashe:2014kda,Baak:2014ora,Eberhardt:2012gv,Ciuchini:2013pca,Ciuchini:2014dea,Charles:2015gya}. The Higgs sector of the SM will be put under similar scrutiny once more data are collected, and even global analyses combining data from all sectors may become possible~\cite{Pomarol:2013zra,Petrov:2015jea}.

While one can examine each new physics model individually against precision data and see what regions of parameter space are allowed (see e.g.~\cite{Erler:2009jh,Heinemeyer:2004gx,Ellis:2007fu,Hubisz:2005tx,Agashe:2005dk,Barbieri:2012tu}), it is often desirable to perform more general analyses whose results can be translated into wide classes of BSM scenarios. In such analyses one usually considers simple extensions of the SM, with a few parameters capturing the leading BSM effects. A well-known example is the $S, T, U$ parameters (or their rescaled versions $\hat S, \hat T, \hat U$), also known as oblique parameters, proposed by Peskin and Takeuchi~\cite{Peskin:1991sw} and later generalized by others~\cite{Maksymyk:1993zm,Barbieri:2004qk}. In fact, thanks to the constraining power of the $Z$-pole data, the oblique parameters formalism has become so influential that it is commonly used for a quick first evaluation of the compatibility of new physics models with data, without enough attention paid to the fact that these parameters are not unambiguously defined in all BSM theories. The problem has become sharper in light of recent efforts to advocate and develop the effective field theory (EFT) framework as the most general (under the assumption of no light new states) model-independent {\it and consistent} approach to precision analyses~\cite{Pomarol:2013zra,Degrande:2012wf,Buchalla:2012qq,Passarino:2012cb,Masso:2012eq,Corbett:2012ja,Grojean:2013kd,Elias-Miro:2013gya,Buchalla:2013wpa,Falkowski:2013dza,Contino:2013kra,Corbett:2013pja,Jenkins:2013fya,Mebane:2013zga,Einhorn:2013kja,Buchalla:2013rka,Elias-Miro:2013mua,Einhorn:2013tja,Jenkins:2013zja,Jenkins:2013wua,Brivio:2013pma,Chen:2013kfa,Alonso:2013hga,Elias-Miro:2013eta,Willenbrock:2014bja,Ellis:2014dva,Belusca-Maito:2014dpa,Alonso:2014zka,Masso:2014xra,Biekoetter:2014jwa,Englert:2014cva,Trott:2014dma,Lehman:2014jma,Ellis:2014jta,Falkowski:2014tna,Henning:2014wua,Gonzalez-Alonso:2014eva,Buchalla:2014eca,Berthier:2015oma,Lehman:2015via,Efrati:2015eaa,Buchalla:2015wfa,Gonzalez-Alonso:2015bha,Falkowski:2015fla,Hartmann:2015oia,Ghezzi:2015vva,Corbett:2015ksa,Chiang:2015ura,Huo:2015exa,Buckley:2015nca,deBlas:2015aea,Wells:2015eba,Bordone:2015nqa,Hartmann:2015aia,Falkowski:2015jaa,Berthier:2015gja,Falkowski:2015wza,Huo:2015nka,Lehman:2015coa,David:2015waa,Brehmer:2015rna,Ellis:2015sca}. Increased interest in this approach has led to different operator bases being proposed, with different motivations, which are all equivalent under field redefinitions~\cite{Grzadkowski:2010es,Contino:2013kra,Elias-Miro:2013mua,Elias-Miro:2013eta}. While the physical {\it observables} are always well-defined independent of the basis choice, naively defining the oblique parameters in the most general EFT is basis-dependent, and is thus not useful.

There are two caveats one should keep in mind when working with the oblique parameters. First, these parameters as defined from the vector boson self-energies $\Pi_{VV'}(p^2)$ are not invariant under redefinitions of the vector boson fields (see~\cite{SanchezColon:1998xg,Grojean:2006nn} for earlier discussions). Thus, unlike observables, they are unphysical and ambiguous unless it is specified how these fields are defined. Second, the bounds on these parameters are usually derived assuming they capture all the BSM effects (or at least the dominant ones) on the processes under study, and so should not be applied to new physics scenarios where this is not the case. In particular, these bounds should not be used to constrain the EFT parameter space, unless restrictions are imposed to satisfy the above assumption.\footnote{The situation is different if measurements of {\it observables} are used to constrain the EFT, in which case no such restrictions are needed. We also note that simultaneously using observables and oblique parameters to constrain the EFT is redundant if such restrictions are imposed, and inconsistent if they are not.} This second caveat actually defines the range of applicability of the oblique parameters analyses, and has been recently emphasized in~\cite{Trott:2014dma}.

The EFT framework as the most general consistent characterization of indirect BSM effects allows these caveats to be properly accounted for. In fact, it is well-known that generally speaking, the usually-quoted constraints on the oblique parameters can be meaningfully interpreted only within universal theories,\footnote{For an extraction of oblique parameters from a particular set of experimental data, the results can also meaningfully constrain some special nonuniversal theories, which are extensions of universal theories by interactions (or effective operators) that do not affect the observables used in this particular extraction, and are thus practically indistinguishable from universal theories without additional experimental information. Aiming at general conclusions, we will not consider this case further in this paper.} where there is a unique well-motivated procedure to eliminate the field-redefinition ambiguity when defining the oblique parameters~\cite{Barbieri:2004qk}. However, a comprehensive EFT description of universal theories is still lacking, and confusion can arise when the oblique parameters are discussed in the EFT context. It is the purpose of this paper to present such a description.

We begin in section~\ref{sec:def} by stating the precise definition of ``universal theories'' in the SMEFT (SM plus the complete set of dimension-6 operators, with linearly-realized electroweak symmetry breaking), both in general terms and in particular operator bases. This will make clear in which cases the oblique parameters analyses can be unambiguously recast in the EFT language, and how the oblique parameters should be written in terms of the Wilson coefficients in each basis. The latter is done in section~\ref{sec:utc}, along with all the other effects universal theories can produce. We will see that universal theories are completely characterized by 16 parameters, dubbed ``universal parameters.'' This number is the same in all SMEFT bases, and the values of the 16 parameters in a particular universal theory are independent of the basis choice. In this framework, the 5 nonvanishing oblique parameters constitute a subset of the 16 universal parameters; the latter also include, e.g.\ the familiar anomalous triple-gauge couplings (TGCs)~\cite{Hagiwara:1986vm} and Higgs coupling rescaling factors~\cite{LHCHiggsCrossSectionWorkingGroup:2012nn}. Next, we connect the universal parameters to the couplings in the Higgs basis~\cite{HiggsBasis} in section~\ref{sec:pheno}. The latter can be directly mapped to new physics corrections to the precision observables, which exhibit a universal pattern. Two examples of corrections to precision observables are discussed in section~\ref{sec:obs}. We recast the calculations of precision electroweak observables in the presence of the most general self-energy corrections in~\cite{Wells:2014pga} in the language of universal parameters (section~\ref{sec:obs-ew}), and demonstrate explicitly the well-known interplay between TGC measurements, especially from $e^+e^-\to W^+W^-$, and Higgs data, in particular the spectrum of the 3-body decay $h\to Z\ell^+\ell^-$ (section~\ref{sec:obs-tgc}). We will see that, despite the concerns raised in~\cite{Trott:2014dma}, it is possible to consistently impose precision electroweak constraints in the form of oblique parameters, and yet find that TGCs and Higgs observables are connected. 
Finally, we conclude in section~\ref{sec:conclusions}. Appendix~\ref{sec:app} collects our notation and some useful formulas.

We will restrict ourselves to leading order in the new physics effects throughout this work. A follow-up paper~\cite{followup} will be devoted to an RG analysis of universal theories.

\section{EFT definition of universal theories}
\label{sec:def}

\subsection{General considerations and bosonic bases}
\label{sec:def-gen}

In the SMEFT with cutoff $\Lambda$, universal theories are defined as theories for which, via field redefinitions, the leading BSM effects can be captured by dimension-6 operators suppressed by $\frac{1}{\Lambda^2}$ which involve SM bosons only (henceforth referred to as ``bosonic operators''). Possible UV completions of such effective theories include not only theories where new states at the scale $\Lambda$ only couple to the bosonic sector of the SM, but also those where the SM fermions are weakly coupled to new states at $\Lambda$ via the vector and/or scalar currents appearing in the SM~\cite{Barbieri:2004qk,Panico:2015jxa}.\footnote{Other possibilities remain. For example, new states much heavier than $\Lambda$ can couple to the SM fermions but not via these currents, since the effective operators generated in that case are suppressed by a much higher scale.} In the latter case, the dimension-6 operators generated involve the SM currents, and can thus be eliminated in favor of bosonic operators via field redefinitions, or equivalently by applying the SM equations of motion (EoM),
\bseq
\beqa
J_{G\mu}^A &\equiv& g_s\sum_{f\in\{q,u,d\}} \bar f \gamma_\mu T^A f \xrightarrow{\EoM} D^\nu G_{\mu\nu}^A, \\
J_{W\mu}^a &\equiv& g\sum_{f\in\{q,l\}} \bar f \gamma_\mu \frac{\sigma^a}{2} f \xrightarrow{\EoM} D^\nu W_{\mu\nu}^a -\frac{ig}{2} H^\dagger\sigma^a\Dlr_\mu H, \\
J_{B\mu} &\equiv& g'\sum_{f\in\{q,l,u,d,e\}} Y_f \bar f \gamma_\mu f \xrightarrow{\EoM} \partial^\nu B_{\mu\nu} -\frac{ig'}{2} H^\dagger\Dlr_\mu H, \\
J_y^\alpha &\equiv& \bar u y_u^\dagger q_\beta \epsilon^{\beta\alpha} + \bar q^\alpha \VCKM y_d d + \bar l^\alpha y_e e \CR
&&\xrightarrow{\EoM} -(D^2 H^\dagger)^\alpha +\lambda v^2 H^{\dagger\alpha} -2\lambda |H|^2 H^{\dagger\alpha},\label{Jydef}
\eeqan
\eseq{Jredef}
where $H^\dagger\sigma^a\Dlr_\mu H = H^\dagger\sigma^a (D_\mu H) - (D_\mu H)^\dagger\sigma^a H$, $H^\dagger\Dlr_\mu H = H^\dagger (D_\mu H) - (D_\mu H)^\dagger H$, $\epsilon^{\beta\alpha}=(i\sigma^2)^{\beta\alpha}$. Here and in the following, all fermions fields are gauge eigenstates unless otherwise specified. $\alpha$, $\beta$ are $SU(2)_L$ indices, while the generation indices are implicitly summed over, with the Yukawa matrices $y_u, y_d, y_e$ diagonal and real in generation space. The latter should not be confused with the hypercharges
\beq
\{Y_q, Y_l, Y_u, Y_d, Y_e\} = \{\frac{1}{6}, -\frac{1}{2}, \frac{2}{3}, -\frac{1}{3}, -1\}.
\eeq{Yf}
The normalizations of the currents have been chosen such that
\beq
\L_\text{SM} \supset G^{A\mu}J_{G\mu}^A + W^{a\mu}J_{W\mu}^a + B^\mu J_{B\mu} - (H_\alpha J_y^\alpha+\text{h.c.}).
\eeqn

There are in total 16 independent CP-even dimension-6 operators one can write down with $D_\mu$ and the SM boson fields $G_{\mu\nu}^A, W_{\mu\nu}^a, B_{\mu\nu}, H$ only. These are enumerated in the first column of table~\ref{tab:olist} above the horizontal solid line, in the notation of~\cite{Elias-Miro:2013mua}. In fact, a redundant set of 18 bosonic operators are listed. There are 2 integration-by-parts (IBP) relations among the 7 operators above the dashed line,
\bseq
\beqa
\O_W &\overset{\IBP}{\longleftrightarrow}& \O_{HW} +\frac{1}{4}(\O_{WW}+\O_{WB}), \\
\O_B &\overset{\IBP}{\longleftrightarrow}& \O_{HB} +\frac{1}{4}(\O_{BB}+\O_{WB}),
\eeqan
\eseq{IBPred}
reducing the set to 16 independent operators. We will neglect the CP-odd operators. With this further restriction, precision flavor physics will not be at play in our discussions, since by definition universal theories satisfy minimal flavor violation (MFV)~\cite{D'Ambrosio:2002ex}. As far as CP-conserving processes in the electroweak and Higgs sectors are concerned, the CP-odd operators only contribute $\O(\frac{v^4}{\Lambda^4})$ corrections and are thus more difficult to probe in general.

\begin{table}[tbp]
\centering
\begin{small}
\begin{tabular}{|c|c|c|c|c|c|}
\hline
Operator & Warsaw & \multicolumn{1}{p{0.45in}|}{\centering EGGM} & \multicolumn{1}{p{0.45in}|}{\centering SILH} & \multicolumn{1}{p{0.25in}|}{\centering \BE} & \multicolumn{1}{p{0.25in}|}{\centering \BS} \\
\hline 
$\O_W=\frac{ig}{2}(H^\dagger\sigma^a\Dlr_\mu H)D^\nu W^a_{\mu\nu}$ & $\times$ &  & & &  \\
$\O_B=\frac{ig'}{2}(H^\dagger\Dlr_\mu H)\partial^\nu B_{\mu\nu}$ & $\times$ &  & & &  \\
$\O_{HW}=ig(D^\mu H)^\dagger\sigma^a(D^\nu H)W^a_{\mu\nu}$ & $\times$ & $\times$ & & $\times$ &  \\
$\O_{HB}=ig'(D^\mu H)^\dagger(D^\nu H)B_{\mu\nu}$ & $\times$ & $\times$ & & $\times$ &  \\
$\O_{WW}=g^2|H|^2W^a_{\mu\nu}W^{a\mu\nu}$ & $Q_{HW}=|H|^2W^a_{\mu\nu}W^{a\mu\nu}$ &  & $\times$ &  & $\times$ \\
$\O_{WB}=gg'H^\dagger\sigma^a HW^a_{\mu\nu}B^{\mu\nu}$ & $Q_{HWB}=H^\dagger\sigma^a HW^a_{\mu\nu}B^{\mu\nu}$ &  & $\times$ &  & $\times$ \\
$\O_{BB}=g'^2|H|^2B_{\mu\nu}B^{\mu\nu}$ & $Q_{HB}=|H|^2B_{\mu\nu}B^{\mu\nu}$ &  & & & \\
\hdashline
$\O_{GG}=g_s^2|H|^2G^A_{\mu\nu}G^{A\mu\nu}$ & $Q_{HG}=|H|^2G^A_{\mu\nu}G^{A\mu\nu}$ &  & &  &  \\
$\O_{2W}=-\frac{1}{2}(D^\mu W^a_{\mu\nu})^2$ & $\times$ & & $\times$ &  & \\
$\O_{2B}=-\frac{1}{2}(\partial^\mu B_{\mu\nu})^2$ & $\times$ & & $\times$ &  & \\
$\O_{2G}=-\frac{1}{2}(D^\mu G^A_{\mu\nu})^2$ & $\times$ & & $\times$ &  & \\
$\O_{3W}=\frac{g}{6}\epsilon^{abc}W_\mu^{a\nu}W_\nu^{b\rho}W_\rho^{c\mu}$ & $Q_W=\epsilon^{abc}W_\mu^{a\nu}W_\nu^{b\rho}W_\rho^{c\mu}$ &  & &  &  \\
$\O_{3G}=\frac{g_s}{6}f^{ABC}G_\mu^{A\nu}G_\nu^{B\rho}G_\rho^{C\mu}$ & $Q_G=f^{ABC}G_\mu^{A\nu}G_\nu^{B\rho}G_\rho^{C\mu}$ &  & &  &  \\
$\O_T=\frac{1}{2}(H^\dagger\Dlr_\mu H)^2$ & $Q_{HD}=|H^\dagger D_\mu H|^2$ & & &  & \\
$\O_H=\frac{1}{2}(\partial_\mu|H|^2)^2$ & $Q_{H\square}=|H|^2\square|H|^2$ &  &  &  & \\
$\O_6=\lambda|H|^6$ & $Q_H=|H|^6$ &  & &  &  \\
$\O_r=|H|^2|D_\mu H|^2$ & $\times$ & $\times$ & $\times$ &  & \\
$\O_{K4}=|D^2H|^2$ & $\times$ & $\times$ & $\times$ &  & \\
\hline
$\O_L^l=(iH^\dagger\Dlr_\mu H)(\bar l\gamma^\mu l)$ & $Q_{Hl}^{(1)}$ &  & $\times$ & \multicolumn{2}{|c|}{} \\
$\O_L^{(3)l}=(iH^\dagger\sigma^a\Dlr_\mu H)(\bar l\gamma^\mu\sigma^a l)$ & $Q_{Hl}^{(3)}$ & $\times$ & $\times$ & \multicolumn{2}{|c|}{} \\
$\O_R^e=(iH^\dagger\Dlr_\mu H)(\bar e\gamma^\mu e)$ & $Q_{He}$ & $\times$ & & \multicolumn{2}{|c|}{}  \\
$\O_{LL}^l=(\bar l\gamma_\mu l)(\bar l\gamma^\mu l)$ & $Q_{ll}$ & $\times$ & & \multicolumn{2}{|c|}{unspecified}  \\
$\O_{RR}^e=(\bar e\gamma_\mu e)(\bar e\gamma^\mu e)$ & $Q_{ee}$ & $\times$ &  & \multicolumn{2}{|c|}{} \\
$\O_{RR}^{(8)ud}=(\bar u\gamma_\mu T^A u)(\bar d\gamma^\mu T^A d)$ & $Q_{ud}^{(8)}$ & $\times$ & & \multicolumn{2}{|c|}{}  \\
\cline{2-4}
other 38 fermionic operators & \multicolumn{3}{|c|}{kept in all 3 bases} & \multicolumn{2}{|c|}{}  \\
\hline
\end{tabular}
\end{small}
\caption{\label{tab:olist} List of CP-even dimension-6 operators (column 1) in the notation of~\cite{Elias-Miro:2013mua}. There are 53 independent operators (for one fermion generation assuming baryon number conservation) among the 24 listed (18 bosonic and 6 fermionic, separated by the horizontal solid line) plus 38 unlisted (fermionic) operators, so 9 of them should be eliminated to form a complete SMEFT basis. The eliminated operators for each of the three recently-proposed bases, Warsaw~\cite{Grzadkowski:2010es}, EGGM~\cite{Elias-Miro:2013eta}, and SILH~\cite{Elias-Miro:2013mua}, are marked by ``$\times$'' (the eliminated fermionic operators refer to the first-generation ones). The operators appear in slightly different forms in the Warsaw basis, where they are denoted by $Q_i$ and are written out explicitly. We also define the \BE\ and \BS\ bases (EGGM-like and SILH-like bosonic bases), each consisting of 16 independent bosonic operators after 2 of the 7 operators above the dashed line are eliminated via IBP. The bosonic bases are complete when describing universal theories at leading order.}
\end{table}

We complete the list of dimension-6 operators by showing those involving SM fermions (henceforth referred to as ``fermionic operators'') below the horizontal solid line in the first column of table~\ref{tab:olist}. It is well-known that the number of independent CP-even dimension-6 operators is 53 (for one fermion generation assuming baryon number conservation). So among the overcomplete set of $18\text{(bosonic)}+6+38\text{(fermionic)}=62$ operators shown in table~\ref{tab:olist}, 9 should be eliminated via field redefinitions to form a complete nonredundant basis. We mark by ``$\times$'' the eliminated operators in each of the 3 recently-proposed SMEFT bases we consider: the Warsaw basis builds upon earlier work~\cite{Buchmuller:1985jz}, and represents the first successful effort to write down a complete nonredundant basis~\cite{Grzadkowski:2010es} (hence it is also known as the standard basis, despite being equivalent to any other basis); the EGGM basis is devised to simplify the study of RG effects in the bosonic sector~\cite{Elias-Miro:2013eta} (see also~\cite{Elias-Miro:2013gya}); the SILH basis originates from the study of the strongly-interacting light Higgs (SILH) scenario~\cite{Giudice:2007fh}, and has been further developed recently~\cite{Contino:2013kra}, resulting in a complete basis being tabulated in~\cite{Elias-Miro:2013mua} under the assumption of MFV. Note that what we refer to as the ``SILH basis'' is the one proposed in~\cite{Elias-Miro:2013mua} in the nonuniversal theories case, and used in the global SMEFT analysis in~\cite{Pomarol:2013zra}. To go beyond MFV, we take the eliminated fermionic operators $\O_L^l$, $\O_L^{(3)l}$ to be those involving the first-generation fermions. The same basis is referred to as the ``SILH$'$ basis'' in~\cite{HiggsBasis}. We have adopted the notation of~\cite{Grzadkowski:2010es} for the Warsaw basis operators $Q_i$ in the second column. For the fermionic operators, $\O_i$ and $Q_i$ differ only by name; for example, $\O_R^e=Q_{He}=(iH^\dagger\Dlr_\mu H)(\bar e\gamma^\mu e)$ represent the same operator. But for the bosonic operators, $\O_i$ and $Q_i$ differ by normalization, so we have written out $Q_i$ explicitly. Furthermore, note that
\beq
\O_T \overset{\IBP}{\longleftrightarrow} -2Q_{HD} -\frac{1}{2}Q_{H\square}
\eeq{OT2Warsaw}
does not directly correspond to $Q_{HD}$, despite the two being in the same row in the table. Also, due to different historical developments of the bases, $Q_{HW}$, $Q_{HB}$, $Q_W$, $Q_H$ are not the same operators as $\O_{HW}$, $\O_{HB}$, $\O_{W}$, $\O_H$; instead, up to normalizations, they correspond to $\O_{WW}$, $\O_{BB}$, $\O_{3W}$, $\O_6$, respectively, as indicated in table~\ref{tab:olist}.

\begin{table}[tbp]
\centering
\begin{tabular}{|p{2in}|p{2in}|}
\hline
$\bar S_W=\bar E_W+4\bar E_{WW}$ & $\bar E_W=\bar S_W+\bar S_{HW}$ \\
$\bar S_B=\bar E_B-4(\bar E_{WW}-\bar E_{WB})$ & $\bar E_B=\bar S_B+\bar S_{HB}$ \\
$\bar S_{HW}=-4\bar E_{WW}$ & $\bar E_{WW}=-\frac{1}{4}\bar S_{HW}$ \\
$\bar S_{HB}=4(\bar E_{WW}-\bar E_{WB})$ & $\bar E_{WB}=-\frac{1}{4}(\bar S_{HW}+\bar S_{HB})$ \\
$\bar S_{BB}=\bar E_{BB}+\bar E_{WW}-\bar E_{WB}$ & $\bar E_{BB}=\bar S_{BB}-\frac{1}{4}\bar S_{HB}$ \\
\hline
\multicolumn{2}{|c|}{$\bar S_i=\bar E_i$ for the other 11 Wilson coefficients.} \\
\hline
\end{tabular}
\caption{\label{tab:BE2BS} Relations between the Wilson coefficients in the \BE\ and \BS\ bases, $\bar E_i$ and $\bar S_i$ in \eqref{LuB}.}
\end{table}

The definition of universal theories stated at the beginning of this subsection can be cast in any complete SMEFT basis. We will discuss this in detail for the 3 recently-proposed bases in the next subsection. However, perhaps the simplest way to completely describe universal theories in the SMEFT is, according to this definition, to use 16 independent CP-even bosonic operators only. We call such a set of 16 bosonic operators a ``bosonic basis,'' in the sense that it can be used as a complete basis for universal theories at leading order. Recall that there is freedom in choosing 5 out of the 7 operators above the dashed line in table~\ref{tab:olist}, and we demonstrate two options -- to eliminate $\O_{HW}$ and $\O_{HB}$, or $\O_{WW}$ and $\O_{WB}$. We call the resulting bosonic bases the EGGM-like and the SILH-like bosonic bases, respectively, or \BE\ and \BS\ bases for short. Denoting the Wilson coefficients in the \BE\ and \BS\ bases by $\bar E_i$ and $\bar S_i$, respectively, we have
\bseq
\beqa
\L_{\text{universal}} &=& \L_{\text{SM}} + \frac{1}{v^2} (\bar E_W\O_W + \bar E_B\O_B + \bar E_{WW}\O_{WW} + \bar E_{WB}\O_{WB} + \bar E_{BB}\O_{BB} \CR
&& + \bar E_{GG}\O_{GG} + \bar E_{2W}\O_{2W} + \bar E_{2B}\O_{2B} + \bar E_{2G}\O_{2G} + \bar E_{3W}\O_{3W} +\bar E_{3G}\O_{3G} \CR
&& + \bar E_T\O_T + \bar E_H\O_H + \bar E_6\O_6 + \bar E_r\O_r + \bar E_{K4}\O_{K4}) \label{LuBE}\\
&=& \L_{\text{SM}} + \frac{1}{v^2} (\bar S_W\O_W + \bar S_B\O_B + \bar S_{HW}\O_{HW} + \bar S_{HB}\O_{HB} + \bar S_{BB}\O_{BB} \CR
&& + \bar S_{GG}\O_{GG} + \bar S_{2W}\O_{2W} + \bar S_{2B}\O_{2B} + \bar S_{2G}\O_{2G} + \bar S_{3W}\O_{3W} +\bar S_{3G}\O_{3G} \CR
&& + \bar S_T\O_T + \bar S_H\O_H + \bar S_6\O_6 + \bar S_r\O_r + \bar S_{K4}\O_{K4}). \label{LuBS}
\eeqan
\eseq{LuB}
The normalization chosen is such that $\bar E_i, \bar S_i\sim\O(\frac{v^2}{\Lambda^2})$. Each of the Wilson coefficient sets $\{\bar E_i\}$ and $\{\bar S_i\}$ spans the 16-dimensional parameter space of universal theories. The translation between the two directly follows from \eqref{IBPred}, and are shown in table~\ref{tab:BE2BS}. Note that while $\bar E_W$, $\bar E_B$, $\bar E_{BB}$ and $\bar S_W$, $\bar S_B$, $\bar S_{BB}$ are the Wilson coefficients of the same three operators, they are not equal numerically and hence have different meanings, because the full sets of operators are not the same in the two bosonic bases.

\subsection{Universal theories in complete SMEFT bases}

In this subsection, we will work out the definition of universal theories in the 3 recently-proposed SMEFT bases, which, unlike the bosonic bases discussed above, are complete and nonredundant for generic nonuniversal theories. In other words, we will find the 16-dimensional subspace of the full SMEFT parameter space that describes universal theories.

\paragraph{EGGM basis.} We start from \eqref{LuBE}, and eliminate $\O_r, \O_{K4}$,
\bseq
\beqa
\O_r &=& |H|^2(D_\mu H)^\dagger(D^\mu H) \xrightarrow{\IBP} -\frac{1}{2} \Bigl[|H|^2 (H^\dagger D^2 H) + (\partial_\mu|H|^2)(H^\dagger D^\mu H) +\text{h.c.}\Bigr] \CR
&=& -\frac{1}{2} |H|^2 (H^\dagger D^2 H +\text{h.c.}) -\O_H \xrightarrow{\EoM} -\lambda v^2 |H|^4 + 2\O_6 + \frac{1}{2} \O_y -\O_H, \label{Or}\\
\O_{K4} &=& |D^2 H|^2 \xrightarrow{\EoM} \CR
&& \lambda^2 v^4 |H|^2 - 4\lambda^2 v^2 |H|^4 + 4\lambda\O_6 - \lambda v^2 (H_\alpha J_y^\alpha+\text{h.c.}) + 2\lambda\O_y + \O_{2y},
\eeqan
\eseq{OrOK4}
where we have defined
\bseq
\beqa
\O_y &\equiv& |H|^2 (H_\alpha J_y^\alpha+\text{h.c.}), \\
\O_{2y} &\equiv& J_{y\alpha}^\dagger J_y^\alpha. \label{O2ydef}
\eeqan
\eseqn
These can be thought of as interactions mediated by a heavy scalar that couples to SM fermions via the scalar current $J_y^\alpha$, i.e.\ in the same way as the SM Higgs field $H$ does. In the EGGM basis, they represent the following linear combinations of operators, with $\O(y_f)$ and $\O(y_f^2)$ coefficients, respectively,
\bseq
\beqa
\O_y &=& [y_u]_{ij} [\O_{y_u}]_{ij} + [\VCKM y_d]_{ij} [\O_{y_d}]_{ij} + [y_e]_{ij} [\O_{y_e}]_{ij} +\text{h.c.}, \label{Oy}\\
\O_{2y} &=& -[y_u]_{il}[y_u^\dagger]_{kj}\Bigl(\frac{1}{6}[\O_{LR}^u]_{ijkl}+[\O_{LR}^{(8)u}]_{ijkl}\Bigr) \CR
&& -[\VCKM y_d]_{il}[y_d^\dagger\VCKM^\dagger]_{kj}\Bigl(\frac{1}{6}[\O_{LR}^d]_{ijkl}+[\O_{LR}^{(8)d}]_{ijkl}\Bigr) -\frac{1}{2}[y_e]_{il}[y_e^\dagger]_{kj}[\O_{LR}^e]_{ijkl} \CR
&& +\Bigl( [y_u]_{ij}[\VCKM y_d]_{kl}[\O_{y_uy_d}]_{ijkl} +[y_u]_{ij}[y_e]_{kl}[\O_{y_uy_e}]_{ijkl} \CR
&& \qquad+[y_e]_{ij}[y_d^\dagger\VCKM^\dagger]_{kl}[\O_{y_ey_d}]_{ijkl} +\text{h.c.} \Bigr).\label{O2y}
\eeqan
\eseq{OyO2y}
Here and in the following, repeated generation indices are summed over unless specified otherwise. Note that our convention slightly differs from that in \cite{Elias-Miro:2013mua} in that we do not include the SM Yukawa couplings in the operators $\O_{y_f}, \O_{y_fy_{f'}}$\footnote{Ref.~\cite{Elias-Miro:2013mua} focuses on one fermion generation when listing the operators. The prescription used there for associating SM Yukawa couplings to operators can be straightforwardly extended to three generations only when MFV is satisfied. We find it useful to factor out the Yukawa couplings, and define universal theories in terms of restrictions on the most general SMEFT that does not assume MFV.}. The appearance of operators involving products of vector currents in \eqref{O2y} is due to Fierz rearrangements, e.g.
\beqa
(\bar q^\alpha y_u u)(\bar u y_u^\dagger q_\alpha) &=& [y_u]_{il}[y_u^\dagger]_{kj} \delta_{ad}\delta_{cb} (\bar q^{a\alpha}_i u^d_l)(\bar u^c_k  q^b_{j\alpha}) \CR
&=& -\frac{1}{2}[y_u]_{il}[y_u^\dagger]_{kj} \delta_{ad}\delta_{cb} (\bar q^{a\alpha}_i \gamma_\mu q^b_{j\alpha})(\bar u^c_k \gamma^\mu u^d_l) \CR
&=& -[y_u]_{il}[y_u^\dagger]_{kj} \Bigl(\frac{1}{6}\delta_{ab}\delta_{cd}+T^A_{ab}T^A_{cd}\Bigr) (\bar q^{a\alpha}_i \gamma_\mu q^b_{j\alpha})(\bar u^c_k \gamma^\mu u^d_l) \CR
&=& -[y_u]_{il}[y_u^\dagger]_{kj} \Bigl[\frac{1}{6}(\bar q_i \gamma_\mu q_j)(\bar u_k \gamma^\mu u_l) + (\bar q_i \gamma_\mu T^A q_j)(\bar u_k \gamma^\mu T^A u_l)\Bigr],
\eeqan
where \eqref{Fierz}, \eqref{groupid} have been used. The generation indices $i,j,k,l$, the $SU(3)_c$ indices $a,b,c,d$, and the $SU(2)_L$ index $\alpha$ have been made explicit where necessary.

The operators with dimensions $\le4$ on the RHS of \eqref{OrOK4} rescale the SM Lagrangian parameters, and have no observable effects. Therefore, in terms of the EGGM basis operators,
\beqa
\L_{\text{universal}} &=& \L_{\text{SM}} + \frac{1}{v^2} (E_W\O_W + E_B\O_B + E_{WW}\O_{WW} + E_{WB}\O_{WB} + E_{BB}\O_{BB} \CR
&& + E_{GG}\O_{GG} + E_{2W}\O_{2W} + E_{2B}\O_{2B} + E_{2G}\O_{2G} + E_{3W}\O_{3W} +E_{3G}\O_{3G} \CR
&& + E_T\O_T + E_H\O_H + E_6\O_6 + E_y\O_y + E_{2y}\O_{2y}).
\eeqa{LuE}
We have denoted the Wilson coefficients by $E_i$ to distinguish from $\bar E_i$ in the \BE\ basis. The translation between $E_i$ and $\bar E_i$ can be read off from \eqref{OrOK4}, and is summarized in table~\ref{tab:BE2E}.

\begin{table}[tbp]
\centering
\begin{tabular}{|p{2in}|p{2in}|}
\hline
$E_H=\bar E_H-\bar E_r$ & $\bar E_H=E_H+2E_y-4\lambda E_{2y}$ \\
$E_6=\bar E_6+2\bar E_r+4\lambda\bar E_{K4}$ & $\bar E_6=E_6-4E_y+4\lambda E_{2y}$ \\
$E_y=\frac{1}{2}\bar E_r+2\lambda\bar E_{K4}$ & $\bar E_r=2E_y-4\lambda E_{2y}$ \\
$E_{2y}=\bar E_{K4}$ & $\bar E_{K4}=E_{2y}$ \\
\hline
\multicolumn{2}{|c|}{$E_i=\bar E_i$ for the other 12 Wilson coefficients.} \\
\hline
\end{tabular}
\caption{\label{tab:BE2E} Relations between the Wilson coefficients in the \BE\ and EGGM bases, $\bar E_i$ in \eqref{LuBE} and $E_i$ in \eqref{LuE}, for universal theories.}
\end{table}

\paragraph{SILH basis.} To translate $\L_{\text{universal}}$ into the SILH basis, we start from \eqref{LuBS}, eliminate $\O_r, \O_{K4}$ as in \eqref{OrOK4}, and further eliminate $\O_{2W}, \O_{2B}, \O_{2G}$ in favor of fermionic operators as follows,
\bseq
\beqa
\O_{2W} &=& -\frac{1}{2} (D^\nu W^a_{\mu\nu})^2 \xrightarrow{\EoM} -\frac{1}{2} (D^\nu W^a_{\mu\nu}) \Bigl(igH^\dagger\sigma^a\Dlr^\mu H -\frac{ig}{2}H^\dagger\sigma^a\Dlr^\mu H +J_W^{a\mu} \Bigr) \CR
&& \xrightarrow{\EoM} -\frac{ig}{2}(H^\dagger\sigma^a\Dlr_\mu H)D^\nu W^a_{\mu\nu} +\frac{1}{2} \Bigl(\frac{ig}{2}H^\dagger\sigma^a\Dlr_\mu H +J^a_{W\mu} \Bigr) \Bigl(\frac{ig}{2}H^\dagger\sigma^a\Dlr^\mu H -J_W^{a\mu} \Bigr) \CR
&& \xrightarrow{\eqref{HsigmaDH}} -\frac{1}{2}g^2\lambda v^2|H|^4 -\O_W -\frac{3}{4}g^2\O_H +g^2\O_6 +\frac{1}{4}g^2\O_y -\frac{1}{2}\O_{2JW}, \label{O2W2S} \\
\O_{2B} &=& -\frac{1}{2} (\partial^\nu B_{\mu\nu})^2 \xrightarrow{\EoM} -\frac{1}{2} (\partial^\nu B_{\mu\nu}) \Bigl(ig'H^\dagger\Dlr^\mu H -\frac{ig'}{2}H^\dagger\Dlr^\mu H +J_B^\mu \Bigr) \CR
&& \xrightarrow{\EoM} -\frac{ig'}{2}(H^\dagger\Dlr_\mu H)\partial^\nu B_{\mu\nu} +\frac{1}{2} \Bigl(\frac{ig'}{2}H^\dagger\Dlr_\mu H +J_{B\mu} \Bigr) \Bigl(\frac{ig'}{2}H^\dagger\Dlr^\mu H -J_B^\mu \Bigr) \CR
&=& -\O_B -\frac{1}{4}g'^2\O_T -\frac{1}{2}\O_{2JB}, \label{O2B2S} \\
\O_{2G} &=& -\frac{1}{2} (D^\nu G^A_{\mu\nu})^2 \xrightarrow{\EoM} -\frac{1}{2}J_{G\mu}^AJ_G^{A\mu} = -\frac{1}{2}\O_{2JG}, \label{O2G2S}
\eeqan
\eseq{O2V}
where we have defined
\bseq
\beqa
\O_{2JW} &\equiv& J_{W\mu}^aJ_W^{a\mu},\\
\O_{2JB} &\equiv& J_{B\mu}J_B^\mu,\\
\O_{2JG} &\equiv& J_{G\mu}^AJ_G^{A\mu}.
\eeqan
\eseq{O2JVdef}
These are linear combinations of SILH basis operators, representing 4-fermion interactions mediated by heavy vector states that couple to the SM vector currents,
\bseq
\beqa
\O_{2JW} &=& g^2\Bigl(-\frac{1}{4}[\O_{LL}^q]_{iijj} +\frac{1}{6}[\O_{LL}^q]_{ijji} +[\O_{LL}^{(8)q}]_{ijji} \CR
&& -\frac{1}{4}[\O_{LL}^l]_{iijj} +\frac{1}{2}[\O_{LL}^l]_{ijji} +\frac{1}{2}[\O_{LL}^{(3)ql}]_{iijj} \Bigr), \\
\O_{2JB} &=& g'^2 \bigl( Y_q^2[\O_{LL}^q]_{iijj} +Y_l^2[\O_{LL}^l]_{iijj} +Y_u^2[\O_{RR}^u]_{iijj} +Y_d^2[\O_{RR}^d]_{iijj} +Y_e^2[\O_{RR}^e]_{iijj} \CR
&& +2Y_qY_l[\O_{LL}^{ql}]_{iijj} +2Y_qY_u[\O_{LR}^u]_{iijj} +2Y_qY_d[\O_{LR}^d]_{iijj} +2Y_qY_e[\O_{LR}^{qe}]_{iijj} \CR
&& +2Y_lY_u[\O_{LR}^{lu}]_{iijj} +2Y_lY_d[\O_{LR}^{ld}]_{iijj} +2Y_lY_e[\O_{LR}^e]_{iijj} \CR
&& +2Y_uY_d[\O_{RR}^{ud}]_{iijj} +2Y_uY_e[\O_{RR}^{ue}]_{iijj} +2Y_dY_e[\O_{RR}^{de}]_{iijj} \bigr),\\
\O_{2JG} &=& g_s^2 \Bigl( [\O_{LL}^{(8)q}]_{iijj} -\frac{1}{6}[\O_{RR}^u]_{iijj} +\frac{1}{2}[\O_{RR}^u]_{ijji} -\frac{1}{6}[\O_{RR}^d]_{iijj} +\frac{1}{2}[\O_{RR}^d]_{ijji} \CR
&& +2[\O_{LR}^{(8)u}]_{iijj} +2[\O_{LR}^{(8)d}]_{iijj} +2[\O_{RR}^{(8)ud}]_{iijj} \Bigr).
\eeqan
\eseqn
Fierz rearrangements \eqref{Fierz} and group-theoretic identities \eqref{groupid} have been used to arrive at the SILH basis operators, e.g.
\bseq
\beqa
(H^\dagger\sigma^a\Dlr_\mu H)^2 &=& \sigma^a_{\alpha\beta}\sigma^a_{\gamma\delta} \Bigl\{\bigl[(H^{\dagger\alpha}D_\mu H^\beta)(H^{\dagger\gamma}D^\mu H^\delta) +\text{h.c.}\bigr] -2 (H^{\dagger\alpha}D_\mu H^\beta)(H^\delta D^\mu H^{\dagger\gamma})\Bigr\} \CR
&=&(2\delta_{\alpha\delta}\delta_{\gamma\beta}-\delta_{\alpha\beta}\delta_{\gamma\delta}) \CR
&&\qquad \Bigl\{\bigl[(H^{\dagger\alpha}D_\mu H^\beta)(H^{\dagger\gamma}D^\mu H^\delta) +\text{h.c.}\bigr] -2 (H^{\dagger\alpha}D_\mu H^\beta)(H^\delta D^\mu H^{\dagger\gamma})\Bigr\} \CR
&=& \bigl[(H^\dagger D_\mu H)^2 +\text{h.c.}\bigr] -4|H|^2|D_\mu H|^2 +2|H^\dagger D_\mu H|^2 \CR
&=& \bigl[H^\dagger (D_\mu H) + (D_\mu H^\dagger) H\bigr]^2 -4|H|^2|D_\mu H|^2 = 2\O_H -4\O_r \CR
&& \xrightarrow{\eqref{Or}} 4\lambda v^2|H|^4 +6\O_H -8\O_6 -2\O_y,
\label{HsigmaDH}\\
(\bar l_i \gamma_\mu \sigma^a l_i)^2 &=& \sigma^a_{\alpha\beta}\sigma^a_{\gamma\delta}(\bar l_i^\alpha \gamma_\mu l_i^\beta)(\bar l_j^\gamma \gamma^\mu l_j^\delta) = (2\delta_{\alpha\delta}\delta_{\gamma\beta}-\delta_{\alpha\beta}\delta_{\gamma\delta})(\bar l_i^\alpha \gamma_\mu l_i^\beta)(\bar l_j^\gamma \gamma^\mu l_j^\delta) \CR
&=& 2\delta_{\alpha\delta}\delta_{\gamma\beta}(\bar l_i^\alpha \gamma_\mu l_j^\delta)(\bar l_j^\gamma \gamma^\mu l_i^\beta) -\delta_{\alpha\beta}\delta_{\gamma\delta}(\bar l_i^\alpha \gamma_\mu l_i^\beta)(\bar l_j^\gamma \gamma^\mu l_j^\delta)  \CR
&=& 2(\bar l_i \gamma_\mu l_j)(\bar l_j \gamma^\mu l_i) -(\bar l_i \gamma_\mu l_i)(\bar l_j \gamma^\mu l_j),
\\
(\bar q_i \gamma_\mu \sigma^a q_i)^2 &=& \delta_{ab}\delta_{cd}(2\delta_{\alpha\delta}\delta_{\gamma\beta}-\delta_{\alpha\beta}\delta_{\gamma\delta})(\bar q_i^{a\alpha} \gamma_\mu q_i^{b\beta})(\bar q_j^{c\gamma} \gamma^\mu q_j^{d\delta}) \CR
&=& 2\delta_{ab}\delta_{cd}\delta_{\alpha\delta}\delta_{\gamma\beta}(\bar q_i^{a\alpha} \gamma_\mu q_j^{d\delta})(\bar q_j^{c\gamma} \gamma^\mu q_i^{b\beta}) -(\bar q_i \gamma_\mu q_i)(\bar q_j \gamma^\mu q_j) \CR
&=& \Bigl(\frac{2}{3}\delta_{ad}\delta_{cb}+4T^A_{ad}T^A_{cb}\Bigr)\delta_{\alpha\delta}\delta_{\gamma\beta}(\bar q_i^{a\alpha} \gamma_\mu q_j^{d\delta})(\bar q_j^{c\gamma} \gamma^\mu q_i^{b\beta}) -(\bar q_i \gamma_\mu q_i)(\bar q_j \gamma^\mu q_j) \CR
&=& \frac{2}{3}(\bar q_i \gamma_\mu q_j)(\bar q_j \gamma^\mu q_i) +4(\bar q_i \gamma_\mu T^A q_j)(\bar q_j \gamma^\mu T^A q_i) -(\bar q_i \gamma_\mu q_i)(\bar q_j \gamma^\mu q_j).
\eeqan
\eseqn

We therefore arrive at the most general Lagrangian for universal theories in terms of the SILH basis operators,
\beqa
\L_{\text{universal}} &=& \L_{\text{SM}} + \frac{1}{v^2} (S_W\O_W + S_B\O_B + S_{HW}\O_{HW} + S_{HB}\O_{HB} + S_{BB}\O_{BB} + S_{GG}\O_{GG} \CR
&& + S_{3W}\O_{3W} +S_{3G}\O_{3G} + S_T\O_T + S_H\O_H + S_6\O_6 \CR
&& + S_{2JW}\O_{2JW} + S_{2JB}\O_{2JB} + S_{2JG}\O_{2JG} + S_y\O_y + S_{2y}\O_{2y}),
\eeqa{LuS}
with the Wilson coefficients denoted by $S_i$, to distinguish from $\bar S_i$ in the \BS\ basis. Note that $\O_y, \O_{2y}$ represent the same operator combinations in the SILH basis as in the EGGM basis, given in \eqref{OyO2y}. The translation between $S_i$ and $\bar S_i$ can be read off from \eqref{OrOK4} and \eqref{O2V}, and is summarized in table~\ref{tab:BS2S}. We can also combine tables~\ref{tab:BE2BS}, \ref{tab:BE2E} and~\ref{tab:BS2S} to derive the dictionary between the EGGM and SILH bases Wilson coefficients in universal theories. This is shown in table~\ref{tab:E2S}. Alternatively, the dictionary can be directly obtained by applying \eqref{O2V} to \eqref{LuE}. 

\begin{table}[tbp]
\centering
\begin{tabular}{|p{2.5in}|p{2.5in}|}
\hline
$S_W=\bar S_W-\bar S_{2W}$ & $\bar S_W=S_W-2S_{2JW}$ \\
$S_B=\bar S_B-\bar S_{2B}$ & $\bar S_B=S_B-2S_{2JB}$ \\
$S_T=\bar S_T-\frac{1}{4}g'^2\bar S_{2B}$ & $\bar S_T=S_T-\frac{1}{2}g'^2S_{2JB}$ \\
$S_H=\bar S_H-\frac{3}{4}g^2\bar S_{2W}-\bar S_r$ & $\bar S_H=S_H-\frac{1}{2}g^2S_{2JW}+2S_y-4\lambda S_{2y}$ \\
$S_6=\bar S_6+g^2\bar S_{2W}+2\bar S_r+4\lambda\bar S_{K4}$ & $\bar S_6=S_6-4S_y+4\lambda S_{2y}$ \\
$S_{2JW}=-\frac{1}{2}\bar S_{2W}$ & $\bar S_{2W}=-2S_{2JW}$ \\
$S_{2JB}=-\frac{1}{2}\bar S_{2B}$ & $\bar S_{2B}=-2S_{2JB}$ \\
$S_{2JG}=-\frac{1}{2}\bar S_{2G}$ & $\bar S_{2G}=-2S_{2JG}$ \\
$S_y=\frac{1}{4}g^2\bar S_{2W}+\frac{1}{2}\bar S_r+2\lambda\bar S_{K4}$ & $\bar S_r=g^2S_{2JW}+2S_y-4\lambda S_{2y}$ \\
$S_{2y}=\bar S_{K4}$ & $\bar S_{K4}=S_{2y}$ \\
\hline
\multicolumn{2}{|c|}{$S_i=\bar S_i$ for the other 6 Wilson coefficients.} \\
\hline
\end{tabular}
\caption{\label{tab:BS2S} Relations between the Wilson coefficients in the \BS\ and SILH bases, $\bar S_i$ in \eqref{LuBS} and $S_i$ in \eqref{LuS}, for universal theories.}
\end{table}

\begin{table}[tbp]
\centering
\begin{tabular}{|p{2.5in}|p{2.5in}|}
\hline
$S_W=E_W+4E_{WW}-E_{2W}$ & $E_W=S_W+S_{HW}-2S_{2JW}$ \\
$S_B=E_B-4(E_{WW}-E_{WB})-E_{2B}$ & $E_B=S_B+S_{HB}-2S_{2JB}$ \\
$S_{HW}=-4E_{WW}$ & $E_{WW}=-\frac{1}{4}S_{HW}$ \\
$S_{HB}=4(E_{WW}-E_{WB})$ & $E_{WB}=-\frac{1}{4}(S_{HW}+S_{HB})$ \\
$S_{BB}=E_{BB}+E_{WW}-E_{WB}$ & $E_{BB}=S_{BB}-\frac{1}{4}S_{HB}$ \\
$S_{GG}=E_{GG}$ & $E_{GG}=S_{GG}$ \\
$S_{3W}=E_{3W}$ & $E_{3W}=S_{3W}$ \\
$S_{3G}=E_{3G}$ & $E_{3G}=S_{3G}$ \\
$S_T=E_T-\frac{1}{4}g'^2E_{2B}$ & $E_T=S_T-\frac{1}{2}g'^2S_{2JB}$ \\
$S_H=E_H-\frac{3}{4}g^2E_{2W}$ & $E_H=S_H-\frac{3}{2}g^2S_{2JW}$ \\
$S_6=E_6+g^2E_{2W}$ & $E_6=S_6+2g^2S_{2JW}$ \\
$S_{2JW}=-\frac{1}{2}E_{2W}$ & $E_{2W}=-2S_{2JW}$ \\
$S_{2JB}=-\frac{1}{2}E_{2B}$ & $E_{2B}=-2S_{2JB}$ \\
$S_{2JG}=-\frac{1}{2}E_{2G}$ & $E_{2G}=-2S_{2JG}$ \\
$S_y=E_y+\frac{1}{4}g^2E_{2W}$ & $E_y=S_y+\frac{1}{2}g^2S_{2JW}$ \\
$S_{2y}=E_{2y}$ & $E_{2y}=S_{2y}$ \\
\hline
\end{tabular}
\caption{\label{tab:E2S} Relations between the Wilson coefficients in the EGGM and SILH bases, $E_i$ in \eqref{LuE} and $S_i$ in \eqref{LuS}, for universal theories.}
\end{table}

\paragraph{Warsaw basis.} Finally, we write $\L_{\text{universal}}$ in the Warsaw basis. Starting from the EGGM basis operators in \eqref{LuE}, we eliminate $\O_W$ and $\O_B$ as follows:
\bseq
\beqa
\O_W &=& \frac{ig}{2}(H^\dagger\sigma^a\Dlr_\mu H)D^\nu W^a_{\mu\nu} \xrightarrow{\EoM} \frac{ig}{2}(H^\dagger\sigma^a\Dlr_\mu H) \Big(\frac{ig}{2} H^\dagger\sigma^a\Dlr^\mu H +J_W^{a\mu}\Bigr) \CR
&& \xrightarrow{\eqref{HsigmaDH}} -g^2\lambda v^2|H|^4 +\frac{3}{4}g^2Q_{H\square} +2g^2\lambda Q_H +\frac{1}{2}g^2Q_y +Q_{HJW}, \label{OW2Warsaw}\\
\O_B &=& \frac{ig'}{2}(H^\dagger\Dlr_\mu H)\partial^\nu B_{\mu\nu} \xrightarrow{\EoM} \frac{ig'}{2}(H^\dagger\Dlr_\mu H) \Bigl(\frac{ig'}{2} H^\dagger\Dlr^\mu H +J_B^\mu\Bigr) \CR
&& \xrightarrow{\eqref{OT2Warsaw}} g'^2Q_{HD} +\frac{1}{4}g'^2Q_{H\square} +Q_{HJB}. \label{OB2Warsaw} 
\eeqan
\eseqn
Here $Q_y\equiv\O_y=|H|^2 (H_\alpha J_y^\alpha+\text{h.c.})$ represent the Warsaw basis operator combination
\beq
Q_y = [y_u]_{ij} [Q_{uH}]_{ij} + [\VCKM y_d]_{ij} [Q_{dH}]_{ij} + [y_e]_{ij} [Q_{eH}]_{ij} +\text{h.c.},
\eeqn
while $Q_{HJW}$ and $Q_{HJB}$ are defined as the following operator combinations in the Warsaw basis,
\bseq
\beqa
Q_{HJW} &\equiv& \frac{ig}{2}(H^\dagger\sigma^a\Dlr_\mu H)J_W^{a\mu} = \frac{1}{4}g^2 \bigl([Q_{Hq}^{(3)}]_{ii} +[Q_{Hl}^{(3)}]_{ii}\bigr), \\
Q_{HJB} &\equiv& \frac{ig'}{2}(H^\dagger\Dlr_\mu H)J_B^\mu \CR
&=& \frac{1}{2}g'^2 \bigl(Y_q[Q_{Hq}^{(1)}]_{ii} +Y_l[Q_{Hl}^{(1)}]_{ii} +Y_u[Q_{Hu}]_{ii} +Y_d[Q_{Hd}]_{ii} +Y_e[Q_{He}]_{ii}\bigr). 
\eeqan
\eseq{QHJV}
In addition to $\O_W$ and $\O_B$, three more operators $\O_{2W}, \O_{2B}, \O_{2G}$ should be eliminated,
\bseq
\beqa
\O_{2W} && \xrightarrow{\eqref{O2W2S}} -\frac{1}{2}g^2\lambda v^2|H|^4 -\O_W -\frac{3}{4}g^2\O_H +g^2\O_6 +\frac{1}{4}g^2\O_y -\frac{1}{2}\O_{2JW} \CR
&& \xrightarrow{\eqref{OW2Warsaw}} \frac{1}{2}g^2\lambda v^2|H|^4 -\frac{3}{8}g^2Q_{H\square} -g^2\lambda Q_H -\frac{1}{4}g^2Q_y -Q_{HJW} -\frac{1}{2}Q_{2JW}, \\
\O_{2B} &&\xrightarrow{\eqref{O2B2S}} -\O_B -\frac{1}{4}g'^2\O_T -\frac{1}{2}\O_{2JB} \CR
&& \xrightarrow{\eqref{OB2Warsaw},\eqref{OT2Warsaw}} -\frac{1}{2}g'^2Q_{HD} -\frac{1}{8}g'^2Q_{H\square} -Q_{HJB} -\frac{1}{2}Q_{2JB}, \\
\O_{2G} && \xrightarrow{\eqref{O2G2S}} -\frac{1}{2}\O_{2JG} =-\frac{1}{2}Q_{2JG},
\eeqan
\eseqn
where $Q_{2JW}, Q_{2JB}, Q_{2JG}$ are the same as $\O_{2JW}, \O_{2JB}, \O_{2JG}$ defined in \eqref{O2JVdef}, but represent linear combinations of Warsaw basis 4-fermion operators,
\bseq
\beqa
Q_{2JW} &=& g^2\Bigl(\frac{1}{4}[Q_{qq}^{(3)}]_{iijj} -\frac{1}{4}[Q_{ll}]_{iijj} +\frac{1}{2}[Q_{ll}]_{ijji} +\frac{1}{2}[Q_{lq}^{(3)}]_{iijj} \Bigr), \\
Q_{2JB} &=& g'^2 \bigl( Y_q^2[Q_{qq}^{(1)}]_{iijj} +Y_l^2[Q_{ll}]_{iijj} +Y_u^2[Q_{uu}]_{iijj} +Y_d^2[Q_{dd}]_{iijj} +Y_e^2[Q_{ee}]_{iijj} \CR
&& +2Y_qY_l[Q_{lq}^{(1)}]_{iijj} +2Y_qY_u[Q_{qu}^{(1)}]_{iijj} +2Y_qY_d[Q_{qd}^{(1)}]_{iijj} +2Y_qY_e[Q_{qe}]_{iijj} \CR
&& +2Y_lY_u[Q_{lu}]_{iijj} +2Y_lY_d[Q_{ld}]_{iijj} +2Y_lY_e[Q_{le}]_{iijj} \CR
&& +2Y_uY_d[Q_{ud}^{(1)}]_{iijj} +2Y_uY_e[Q_{eu}]_{iijj} +2Y_dY_e[Q_{ed}]_{iijj} \bigr), \\
Q_{2JG} &=&  g_s^2 \Bigl( -\frac{1}{6}[Q_{qq}^{(1)}]_{iijj} +\frac{1}{4}[Q_{qq}^{(1)}]_{ijji} +\frac{1}{4}[Q_{qq}^{(3)}]_{ijji} -\frac{1}{6}[Q_{uu}]_{iijj} +\frac{1}{2}[Q_{uu}]_{ijji} \CR
&& -\frac{1}{6}[Q_{dd}]_{iijj} +\frac{1}{2}[Q_{dd}]_{ijji} +2[Q_{qu}^{(8)}]_{iijj} +2[Q_{qd}^{(8)}]_{iijj} +2[Q_{ud}^{(8)}]_{iijj} \Bigr).
\eeqan
\eseq{Q2JV}
Similarly, we use $Q_{2y}$ to denote the combination corresponding to $\O_{2y}$ defined in \eqref{O2ydef},
\beqa
Q_{2y} &=& -[y_u]_{il}[y_u^\dagger]_{kj}\Bigl(\frac{1}{6}[Q_{qu}^{(1)}]_{ijkl}+[Q_{qu}^{(8)}]_{ijkl}\Bigr) \CR
&& -[\VCKM y_d]_{il}[y_d^\dagger\VCKM^\dagger]_{kj}\Bigl(\frac{1}{6}[Q_{qd}^{(1)}]_{ijkl}+[Q_{qd}^{(8)}]_{ijkl}\Bigr) -\frac{1}{2}[y_e]_{il}[y_e^\dagger]_{kj}[Q_{le}]_{ijkl} \CR
&& +\Bigl( [y_u]_{ij}[\VCKM y_d]_{kl}[Q_{quqd}^{(1)}]_{ijkl} -[y_e]_{ij}[y_u]_{kl}[Q_{lequ}^{(1)}]_{ijkl} \CR
&& \qquad+[y_e]_{ij}[y_d^\dagger\VCKM^\dagger]_{kl}[Q_{ledq}]_{ijkl} +\text{h.c.} \Bigr).
\eeqan

\begin{table}[tbp]
\centering
\begin{tabular}{|p{2.75in}|p{2.75in}|}
\hline
$C_{HW}=g^2E_{WW}$ & $E_{WW}=\frac{1}{g^2}C_{HW}$ \\
$C_{HWB}=gg'E_{WB}$ & $E_{WB}=\frac{1}{gg'}C_{HWB}$ \\
$C_{HB}=g'^2E_{BB}$ & $E_{BB}=\frac{1}{g'^2}C_{HB}$ \\
$C_{HG}=g_s^2E_{GG}$ & $E_{GG}=\frac{1}{g_s^2}C_{HG}$ \\
$C_W=\frac{1}{6}gE_{3W}$ & $E_{3W}=\frac{6}{g}C_W$ \\
$C_G=\frac{1}{6}g_sE_{3G}$ & $E_{3G}=\frac{6}{g_s}C_G$ \\
$C_{HD}=-2E_T+\frac{1}{2}g'^2(2E_B-E_{2B})$ & $E_T=-\frac{1}{2}C_{HD}+\frac{1}{2}g'^2(C_{HJB}-C_{2JB})$ \\
$C_{H\square}=-\frac{1}{2}E_H-\frac{1}{2}E_T+\frac{1}{8}g'^2(2E_B-E_{2B})$ & $E_H=-2C_{H\square}+\frac{1}{2}C_{HD}$ \\
$\qquad\quad\,\,\, +\frac{3}{8}g^2(2E_W-E_{2W})$ & $\qquad\quad +\frac{3}{2}g^2(C_{HJW}-C_{2JW})$ \\
$C_H=\lambda E_6+g^2\lambda(2E_W-E_{2W})$ & $E_6=\frac{1}{\lambda}C_H-2g^2(C_{HJW}-C_{2JW})$ \\
$C_{HJW}=E_W-E_{2W}$ & $E_W=C_{HJW}-2C_{2JW}$ \\
$C_{HJB}=E_B-E_{2B}$ & $E_B=C_{HJB}-2C_{2JB}$ \\
$C_{2JW}=-\frac{1}{2}E_{2W}$ & $E_{2W}=-2C_{2JW}$ \\
$C_{2JB}=-\frac{1}{2}E_{2B}$ & $E_{2B}=-2C_{2JB}$ \\
$C_{2JG}=-\frac{1}{2}E_{2G}$ & $E_{2G}=-2C_{2JG}$ \\
$C_y=E_y+\frac{1}{4}g^2(2E_W-E_{2W})$ & $E_y=C_y-\frac{1}{2}g^2(C_{HJW}-C_{2JW})$ \\
$C_{2y}=E_{2y}$ & $E_{2y}=C_{2y}$ \\
\hline
\end{tabular}
\caption{\label{tab:E2W} Relations between the Wilson coefficients in the EGGM and Warsaw bases, $E_i$ in \eqref{LuE} and $C_i$ in \eqref{LuW}, for universal theories.}
\end{table}

Following the procedures detailed above, we obtain the universal theories Lagrangian in terms of the Warsaw basis operators,
\beqa
\L_{\text{universal}} &=& \L_{\text{SM}} + \frac{1}{v^2} (C_{HW}Q_{HW} + C_{HWB}Q_{HWB} + C_{HB}Q_{HB} + C_{HG}Q_{HG} \CR
&& + C_{W}Q_{W} +C_{G}Q_{G} + C_{HD}Q_{HD} + C_{H\square}Q_{H\square} + C_HQ_H \CR
&& +C_{HJW}Q_{HJW} +C_{HJB}Q_{HJB} + C_{2JW}Q_{2JW} + C_{2JB}Q_{2JB} + C_{2JG}Q_{2JG} \CR
&& + C_yQ_y + C_{2y}Q_{2y}),
\eeqa{LuW}
with the Wilson coefficients denoted by $C_i$ (instead of $W_i$ to avoid clash of notation with the $W$'s in the subscripts). They are related to EGGM basis coefficients $E_i$ by the basis transformation summarized in table~\ref{tab:E2W}.

\bigskip
To sum up, eqs.~\eqref{LuE}, \eqref{LuS} and \eqref{LuW} represent the definition of universal theories in the EGGM, SILH, and Warsaw bases, respectively, with Wilson coefficients related to the bosonic bases and to each other as shown in tables~\ref{tab:BE2E}, \ref{tab:BS2S}, \ref{tab:E2S} and~\ref{tab:E2W}. Eqs.~\eqref{LuB}, \eqref{LuE}, \eqref{LuS} and \eqref{LuW} are equivalent effective Lagrangians at the dimension-6 level, and can be transformed into each other via field redefinitions. Independent of the basis choice, there are always 16 independent Wilson coefficients in $\L_{\text{universal}}$. We emphasize that {\it this is the number of independent bosonic operators one can possibly write down, rather than the number of bosonic operators in any particular basis}. In fact, the latter number is 16, 14, 11, and 9 in the bosonic (\BE\ and \BS), EGGM, SILH, and Warsaw bases, respectively, as is clear from table~\ref{tab:olist}. In each of the 3 complete bases discussed in this subsection, there are (combinations of) fermionic operators that are part of $\L_{\text{universal}}$, and should not be discarded for a consistent analysis of universal theories aiming at basis-independent conclusions.\footnote{It is claimed in section~2 of~\cite{Elias-Miro:2013mua} that the number of independent parameters in universal theories is 14. This is because $\O_y$ and $\O_{2y}$ are left out in the counting. However, the presence of $\O_y$ in universal theories is recognized in section~6 of that paper.}

As a side remark, it is often argued (see e.g.~\cite{Elias-Miro:2013gya,Elias-Miro:2013mua}) that the Warsaw basis is less convenient for studying universal theories, because new physics effects are encoded in the correlations among various Wilson coefficients of the fermionic operators; see e.g.~\eqref{QHJV}, \eqref{Q2JV} above. While this is true in many cases, the Warsaw basis does have the capability of describing universal theories as well as any other basis. In fact, the form of $\L_{\text{universal}}$ in the Warsaw basis that we have worked out will be useful in the discussion of RG effects in~\cite{followup}, since the full anomalous dimension matrix for the dimension-6 operators has only been calculated in this basis.

\section{Characterization of universal theories: oblique parameters and beyond}
\label{sec:utc}

In this section, we present an unambiguous and basis-independent definition of the oblique parameters in universal theories,\footnote{By ``basis-independent,'' we mean that the values of the oblique parameters (and more generally universal parameters to be defined below), as calculated in the SMEFT, are the same for a particular universal theory, no matter what basis of SMEFT it is matched onto.} and further develop a formalism for the characterization of universal theories that generalizes the oblique parameters framework. In particular, we transform $\L_{\text{universal}}$ via field and parameter redefinitions into a form where coefficients of various terms are identified with what we call universal parameters, a set of 16 independent parameters that completely characterizes universal theories. It is convenient to first work with the EGGM basis. We will later translate the results into other bases with the help of the dictionaries worked out in the previous section. To make the physics transparent, we write $\L_{\text{universal}}$ in the unitary gauge,
\beq
\L_{\text{universal}} = \L_{V^2} +\L_{V^3} +\L_h +\L_{hV} +\L_{hf} +\L_{4f} +\L_{fDf} +\O(V^4).
\eeqn
The various terms are:
\begin{itemize}
\item Gauge boson quadratic terms
\beqa
\L_{V^2} &=& \Bigl(\frac{gv}{2}\Bigr)^2 W^+_\mu W^{-\mu} +(1-E_T) \frac{1}{2} \Bigl(\frac{gv}{2\cw}\Bigr)^2 Z_\mu Z^\mu \CR
&& -(1-2g_s^2 E_{GG}) \frac{1}{2}G^A_\mu \hat K^{\mu\nu} G^A_\nu \CR
&&\quad -\biggl[1-g^2\Bigl(\frac{1}{2}E_W+2E_{WW}\Bigr)\biggr] \Bigl(W^+_\mu \hat K^{\mu\nu} W^-_\nu +\frac{1}{2} W^3_\mu \hat K^{\mu\nu} W^3_\nu \Bigr) \CR
&&\quad -gg' \Bigl(E_{WB}+\frac{1}{4}E_W+\frac{1}{4}E_B\Bigr) W^3_\mu \hat K^{\mu\nu} B_\nu \CR
&&\quad -\biggl[1-g'^2\Bigl(\frac{1}{2}E_B+2E_{BB}\Bigr)\biggr] \frac{1}{2}B_\mu \hat K^{\mu\nu} B_\nu \CR
&& -\frac{1}{v^2}\biggl[E_{2G} \frac{1}{2} G^A_\mu \hat K^{2\mu\nu} G^A_\nu \CR
&& \qquad +E_{2W} \Bigl(W^+_\mu \hat K^{2\mu\nu} W^-_\nu +\frac{1}{2} W^3_\mu \hat K^{2\mu\nu} W^3_\nu\Bigr) +E_{2B} \frac{1}{2} B_\mu \hat K^{2\mu\nu} B_\nu \Bigr],
\eeqa{LV2}
where $\cw$ is short for $\cos\theta_w=\frac{g}{\sqrt{g^2+g'^2}}$ (similarly we will denote $\sin\theta_w$ by $\sw$), and
\beq
\hat K^{\mu\nu} \equiv -g^{\mu\nu}\partial^2+\partial^\mu\partial^\nu, \quad \hat K^{2\mu\nu} \equiv \hat K^{\mu\rho}\hat K_\rho^{\,\,\,\nu}.
\eeq{Kdef}
\item Triple-gauge interactions
\beqa
\L_{V^3} &=& ig \biggl\{ (W^+_{\mu\nu}W^{-\mu}-W^-_{\mu\nu}W^{+\mu}) \biggl[ \Bigl( 1-\frac{g^2}{2}\bigl(1+\frac{1}{2\cw^2}\bigr)E_W -2g^2E_{WW} \Bigr) \cw Z^\nu \CR
&&\qquad +\Bigl(1 -\frac{g^2}{2}E_W -2g^2E_{WW}\Bigr) \sw A^\nu \biggr] \CR
&& +\frac{1}{2}W^+_{[\mu,}W^-_{\nu]} \biggl[ \Bigl(1-g'^2E_{WB} -\frac{g^2}{2}\bigl(1+\frac{1}{2\cw^2}\bigr)E_W -2g^2E_{WW}\Bigr) \cw Z^{\mu\nu} \CR
&&\qquad +\Bigl(1 +g^2E_{WB} -\frac{g^2}{2}E_W -2g^2E_{WW}\Bigr) \sw A^{\mu\nu} \biggr] \CR
&& -\frac{E_{3W}}{v^2} W^{+\nu}_\mu W^{-\rho}_\nu (\cw{Z_\rho}^\mu+\sw{A_\rho}^\mu) \biggr\} +\frac{E_{2W}}{v^2}\hat K\circ \L_{WWV}^{\SM} \CR
&& +(1-2g_s^2E_{GG})\L_{G^3}^{\SM} +\frac{E_{3G}}{v^2}\frac{g_s}{6}f^{ABC}G_\mu^{A\nu}G_\nu^{B\rho}G_\rho^{C\mu} +\frac{E_{2G}}{v^2}\hat K\circ \L_{G^3}^{\SM},
\eeqa{LV3}
where $W^+_{[\mu,}W^-_{\nu]}=W^+_\mu W^-_\nu-W^+_\nu W^-_\mu$, $W^\pm_{\mu\nu}=\partial_{[\mu,}W^\pm_{\nu]}$. $\L_{WWV}^{\SM}$ can be read off from the terms in the curly bracket in above equation by setting $E_i\to0$. The action of $\hat K\circ$ follows the product rule, e.g.
\beqa
&&\hat K\circ (W^+_{\mu\nu}W^{-\mu}Z^\nu) = \hat K\circ (\partial_{[\mu,}W^+_{\nu]}W^{-\mu}Z^\nu) \CR
&&\qquad =\partial_{[\mu,}(\hat K W^+)_{\nu]} W^{-\mu}Z^\nu +\partial_{[\mu,}W^+_{\nu]}(\hat K W^-)^\mu Z^\nu +\partial_{[\mu,}W^+_{\nu]}W^{-\mu}(\hat K Z)^\nu,
\eeqa{Kcirc}
where $(\hat K W^+)_{\nu}=\hat K_{\nu\rho} W^{+\rho}$, etc. For the special case of $\bar ff\to W^+W^-$ at tree level, assuming on-shell $W^+W^-$ and $m_f=0$, 
\beq
\hat K^{\mu\nu}\to-g^{\mu\nu}\partial^2 \to
\begin{cases}
g^{\mu\nu}m_W^2 &\text{for outgoing}\,\, W^\pm, \\
g^{\mu\nu}\hat s &\text{for $s$-channel}\,\, Z^*/\gamma^*,
\end{cases}
\eeqn
where $\hat s$ is the partonic center-of-mass energy squared. The effect of $\hat K\circ \L_{WWV}^{\SM}$ is thus equivalent to $(\hat s+2m_W^2)\L_{WWV}^{\SM}$ in momentum space in this case.
\item Higgs boson kinetic and potential terms
\beqa
\L_{h} &=& \biggl[1+E_H\Bigl(1+\frac{h}{v}\Bigr)^2\biggr] \frac{1}{2}\partial_\mu h \partial^\mu h \CR
&& -\Bigl(1-\frac{3}{2}E_6\Bigr) \frac{1}{2}(2\lambda v^2) h^2 -\Bigl(1-\frac{5}{2}E_6\Bigr) \lambda v h^3 +\O(h^4).
\eeqa{Lh}
Note that due to the presence of $\O_6=\lambda|H|^4$, the Higgs potential has been modified such that the original parameter $v$ in the SM Lagrangian no longer represents the minimum of the potential. In fact, the minimum $|\langle H\rangle|$ has shifted from $\frac{v}{\sqrt{2}}$ to $(1+\frac{3}{8}E_6)\frac{v}{\sqrt{2}}$. Therefore, we have redefined $(1+\frac{3}{8}E_6)v\to v$, so that the parameter $v$ in \eqref{Lh} represents the true minimum, and is thus the same $v$ that appears in all other parts of $\L_{\text{universal}}$ derived by expanding $H$ around the true minimum $H=\frac{1}{\sqrt{2}}(0,v+h)$.
\item Higgs-fermion interactions
\beq
\L_{hf} = -\Bigl[1 +(1-E_y)\frac{h}{v} -\frac{3}{2}E_y\frac{h^2}{v^2}\Bigr] \Bigl(1-\frac{1}{2}E_y\Bigr) \sum_{f'} \frac{y_{f'} v}{\sqrt{2}} \bar f' f' +\O(h^3f^2),
\eeq{Lhf}
where the sum is over mass eigenstates, denoted by $f'$ to distinguish from the gauge eigenstates $f$.
\item Higgs-vector boson interactions
\beqa
\L_{hV} &=& 2\Bigl(\frac{h}{v}+\frac{h^2}{2v^2}\Bigr) \Bigl(\frac{gv}{2}\Bigr)^2 W^+_\mu W^{-\mu} \CR
&&\qquad +\Bigl[(1-2E_T)\frac{h}{v} +(1-6E_T)\frac{h^2}{2v^2}\Bigr] \Bigl(\frac{gv}{2\cw}\Bigr)^2 Z_\mu Z^\mu \CR
&& +\Bigl(\frac{h}{v}+\frac{h^2}{2v^2}\Bigr) \Bigl\{ E_{GG}g_s^2 \partial_{[\mu,}G^A_{\nu]} \partial^{[\mu,}G^{A\nu]} +2E_{WW}g^2 W^+_{\mu\nu}W^{-\mu\nu} \CR
&&\qquad +(\cw^4E_{WW}+\cw^2\sw^2E_{WB}+\sw^4E_{BB})\frac{g^2}{\cw^2}Z_{\mu\nu}Z^{\mu\nu} \CR
&&\qquad +\bigl[2\cw^2E_{WW}-(\cw^2-\sw^2)E_{WB}-2\sw^2E_{BB}\bigr]gg'Z_{\mu\nu}A^{\mu\nu} \CR
&&\qquad +(E_{WW}-E_{WB}+E_{BB})e^2 A_{\mu\nu}A^{\mu\nu} \CR
&&\qquad +\frac{1}{2} \bigl[E_Wg^2(W^-_\mu\partial_\nu W^{+\mu\nu}+\text{h.c.}) + (E_Wg^2+E_Bg'^2)Z_\mu\partial_\nu Z^{\mu\nu} \CR
&&\qquad\qquad +(E_W-E_B)gg'Z_\mu\partial_\nu A^{\mu\nu}\bigr]\Bigr\} +\O(hV^3, h^3V^2).
\eeqa{LhV}
\item Four-fermion interactions
\beq
\L_{4f} = E_{2y}\O_{2y},
\eeq{L4f}
with $\O_{2y}$ given in \eqref{O2y}.
\item Gauged fermion kinetic terms (same as in the SM)
\beq
\L_{fDf} = \sum_{f\in\{q,l,u,d,e\}} i\bar f\gamma^\mu D_\mu f.
\eeq{LfDf}
\end{itemize}
In all the equations above, the fields and parameters are the SM ones, with the exception of the parameter $v$. No field or parameter redefinitions have been made except for the rescaling of $v$ (and the associated redefinition of $h$) explained below \eqref{Lh}.

\subsection{Oblique parameters}
\label{sec:utc-obl}

In universal theories, the oblique parameters are defined from the Taylor expansion coefficients of the new physics contributions to the transverse part of the vector boson self-energies $\Pi_{VV'}(p^2)$ (defined with the SM piece excluded), 
\beqa
&& \Pi_{VV'}^{\mu\nu}(p^2) = \Bigl(g^{\mu\nu}-\frac{p^\mu p^\nu}{p^2}\Bigr)\Pi_{VV'}(p^2) + \frac{p^\mu p^\nu}{p^2}(\dots) \CR
&& \text{where}\quad \Pi_{VV'}(p^2) = \Pi_{VV'}(0) + \Pi_{VV'}'(0)p^2 + \frac{1}{2}\Pi_{VV'}''(0)(p^2)^2 +\dots,
\eeqa{Pidef}
with the vector boson fields and the SM parameters redefined such that the following 3 {\it oblique parameters defining conditions} are satisfied~\cite{Barbieri:2004qk}:
\begin{itemize}
\item[1)] Only bosonic operators are present.
\item[2)] The kinetic terms of $W^\pm$ and $B$ are canonically normalized.
\item[3)] $\Pi_{WW}(0)=0$ [here $W$ represents $W^\pm$, see \eqref{LV2Pi} below].
\end{itemize}
In particular, the nonzero oblique parameters 
in the linear SMEFT up to dimension 6 are defined by
\bseq
\beqa
\hat S &\equiv& \frac{\alpha}{4\sw^2}S \equiv -\frac{\cw}{\sw}\bar\Pi_{3B}'(0), \\
\hat T &\equiv& \alpha T \equiv \frac{1}{m_W^2} \bigl[\bar\Pi_{WW}(0)-\bar\Pi_{33}(0)\bigr], \\
W &\equiv& -\frac{m_W^2}{2} \bar\Pi_{33}''(0), \\
Y &\equiv& -\frac{m_W^2}{2} \bar\Pi_{BB}''(0) \\
Z &\equiv& -\frac{m_W^2}{2} \bar\Pi_{GG}''(0),
\eeqan
\eseq{obldef}
where $\bar\Pi_{VV'}$ are the self-energies of the vector boson fields after redefinitions are performed (to be explicitly shown below) to satisfy the 3 oblique parameters defining conditions stated above. In these equations one can use the SM leading-order expressions for $m_W$, and $\cw,\sw$ before the redefinitions, since $\bar\Pi_{VV'}$ are already $\O(\frac{v^2}{\Lambda^2})$; the same applies to all the $\O(\frac{v^2}{\Lambda^2})$ terms and will be implicitly understood in various equations in the following. Our sign conventions differ from~\cite{Barbieri:2004qk} but agree with the commonly-used ones. Note that the $U$ parameter (or its rescaled version $\hat U$) originally defined in~\cite{Peskin:1991sw} is zero at the dimension-6 level.

The definitions of oblique parameters are unambiguous from the 3 defining conditions stated above: the first condition dictates the use of a bosonic basis; the second and third conditions fix the SM parameters $g, g', v$ so that there is no more freedom to rescale them within the bosonic basis. In a sense, the intrinsic ambiguity of defining oblique parameters from self-energies is eliminated by choosing a well-motivated prescription for field redefinitions, namely to eliminate all fermionic operators and go to a bosonic basis. The latter is possible only in universal theories. Once the choice is made, no further field redefinitions are allowed since they will reintroduce the currents containing SM fermions and hence fermionic operators.  In nonuniversal theories, on the other hand, precision analyses with oblique parameters are in general inappropriate (and observables should be used instead), since it is not possible to shuffle all the leading BSM effects into the bosonic sector, as required by the oblique parameters defining conditions. In particular, any attempt to define oblique parameters from $\Pi_{VV'}(p^2)$ in the general SMEFT, where all dimension-6 operators are present, is dependent on the choice of basis, i.e.\ on which fermionic operators are kept in the basis, because the latter determines the meaning of the Wilson coefficients contributing to $\Pi_{VV'}(p^2)$. Transforming from one basis to another generally changes the values of bosonic operator Wilson coefficients, and hence the values of $\Pi_{VV'}(p^2)$. Thus, it is impossible to derive basis-independent constraints on nonuniversal theories from the bounds on the oblique parameters naively defined from self-energy corrections -- the procedure is not consistent since the full SMEFT parameter space is much larger than bosonic operators alone can span.

In passing, however, we remark that in some special cases, an analysis with oblique parameters supplemented by additional anomalous fermion couplings can be appropriate and useful. For example, in theories where the heavy states couple preferentially to the third-generation SM fermions, it may be possible to redefine the fields and parameters such that the leading BSM effects in the electroweak sector are completely characterized by the oblique parameters plus anomalous third-generation fermion couplings.\footnote{This case is particularly interesting also from the RG point of view, because the third-generation couplings receive larger loop corrections proportional to $y_f^2$; see~\cite{followup}.} In this case, one can meaningfully talk about constraints on the oblique parameters despite the theory being nonuniversal, but should nevertheless keep in mind that they should be derived from a fit including the anomalous third-generation fermion couplings also; see e.g.~\cite{Gori:2015nqa} for a recent analysis. The often-quoted constraints on the $S, T$ parameters assume the absence of such anomalous fermion couplings, and thus should not be applied to this case.

Now we make the arguments above concrete, by deriving the 5 oblique parameters in terms of the SMEFT Wilson coefficients. In principle, we should work with a bosonic basis, e.g.\ the \BE\ basis, instead of the EGGM basis to satisfy the first oblique parameters defining condition stated above. But in practice, for all the Wilson coefficients that appear in this derivation, $\bar E_i=E_i$, so we will omit the bars for simplicity and write $E_i$ instead of $\bar E_i$. To begin with, \eqref{LV2} can be rewritten as follows,
\beqa
\L_{V^2} &=& \biggl[\Bigl(\frac{gv}{2}\Bigr)^2 +\Pi_{WW}(0)\biggr] W^+_\mu W^{-\mu} +\biggl[\Bigl(\frac{gv}{2\cw}\Bigr)^2+\Pi_{ZZ}(0)\biggr] \frac{1}{2} Z_\mu Z^\mu \CR
&& -\bigl[1-\Pi_{GG}'(0)\bigr] \frac{1}{2}G^A_\mu \hat K^{\mu\nu} G^A_\nu -\bigl[1-\Pi_{WW}'(0)\bigr] W^+_\mu \hat K^{\mu\nu} W^-_\nu \CR
&& -\bigl[1-\Pi_{33}'(0)\bigr] \frac{1}{2}W^3_\mu \hat K^{\mu\nu} W^3_\nu +\Pi_{3B}'(0) W^3_\mu \hat K^{\mu\nu} B_\nu -\bigl[1-\Pi_{BB}'(0)\bigr] \frac{1}{2}B_\mu \hat K^{\mu\nu} B_\nu \CR
&& \CR
&& +\frac{1}{2}\Bigl[\Pi_{GG}''(0) \frac{1}{2}G^A_\mu \hat K^{2\mu\nu} G^A_\nu +\Pi_{WW}''(0) W^+_\mu \hat K^{2\mu\nu} W^-_\nu +\Pi_{33}''(0) \frac{1}{2}W^3_\mu \hat K^{2\mu\nu} W^3_\nu \CR
&& \qquad +\Pi_{3B}''(0) W^3_\mu \hat K^{2\mu\nu} B_\nu +\Pi_{BB}''(0) \frac{1}{2}B_\mu \hat K^{2\mu\nu} B_\nu \Bigr].
\eeqa{LV2Pi}
One can easily get the Taylor expansion coefficients of $\Pi_{VV'}(p^2)$ in the EGGM basis by comparing \eqref{LV2Pi} with \eqref{LV2},
\bseq
\beqa
&& \Pi_{WW}(0) = 0,\quad \Pi_{ZZ}(0) = -m_Z^2E_T, \label{PiE0}\\
&& \Pi_{GG}'(0) = 2g_s^2E_{GG},\quad \Pi_{WW}'(0) = \Pi_{33}'(0) = g^2\Bigl(\frac{1}{2}E_W+2E_{WW}\Bigr), \CR
&& \Pi_{3B}'(0) = -gg'\Bigl(E_{WB}+\frac{1}{4}E_W+\frac{1}{4}E_B\Bigr),\quad \Pi_{BB}'(0) = g'^2\Bigl(\frac{1}{2}E_B+2E_{BB}\Bigr), \\
&& \Pi_{GG}''(0) = -\frac{2}{v^2}E_{2G},\quad \Pi_{WW}''(0) = \Pi_{33}''(0) = -\frac{2}{v^2}E_{2W},\CR
&& \Pi_{3B}''(0) = 0,\quad \Pi_{BB}''(0) = -\frac{2}{v^2}E_{2B}.\quad
\eeqan
\eseq{PiE}
We then carry out the following field and parameter redefinitions,
\bseq
\beqa
&& g_s = \Bigl[1 -\frac{1}{2}\Pi_{GG}'(0)\Bigr] \bar g_s,\quad G^A_\mu = \Bigl[1 +\frac{1}{2}\Pi_{GG}'(0)\Bigr] \bar G^A_\mu,\\
&& g = \Bigl[1 -\frac{1}{2}\Pi_{WW}'(0)\Bigr] \bar g,\quad W^a_\mu = \Bigl[1 +\frac{1}{2}\Pi_{WW}'(0)\Bigr] \bar W^a_\mu,\\
&& g' = \Bigl[1 -\frac{1}{2}\Pi_{BB}'(0)\Bigr] \bar g',\quad B_\mu = \Bigl[1 +\frac{1}{2}\Pi_{BB}'(0)\Bigr] \bar B_\mu,\\
&& v = \Bigl[1 -\frac{1}{2}\frac{\Pi_{WW}(0)}{m_W^2}\Bigr] \bar v.\label{vredef}
\eeqan
\eseq{STWYredef}
These redefinitions make the kinetic terms for $\bar G^A_\mu, \bar W^\pm_\mu, \bar B_\mu$ canonical so as to satisfy the second oblique parameters defining condition, and meanwhile ensure $g_sG^A_\mu=\bar g_s\bar G^A_\mu, gW^a_\mu=\bar g\bar W^a_\mu, g'B_\mu=\bar g'\bar B_\mu$, so that all gauge interactions of the matter fields (SM fermions and Higgs) preserve their SM forms. In other words, no fermionic dimension-6 operators are generated and the first oblique parameters defining condition is still satisfied. The redefinition of $v$ is not really necessary in the \BE\ basis since $\Pi_{WW}(0)$ is already zero (the third oblique parameters defining condition is already satisfied), but we will keep the calculation more general in this subsection. Similarly, while $\Pi_{WW}'(0)=\Pi_{33}'(0)$, $\Pi_{WW}''(0)=\Pi_{33}''(0)$, $\Pi_{3B}''(0)=0$ at the dimension-6 level (corresponding to the additional oblique parameters $\hat U, V, X$~\cite{Peskin:1991sw,Maksymyk:1993zm,Barbieri:2004qk} being zero), 
we have kept separately all 5 parameters for generality. From \eqref{STWYredef} we also have
\bseq
\beqa
\cw &=& \frac{g}{\sqrt{g^2+g'^2}} = \Bigl[1 -\frac{\sw^2}{2}\Pi_{WW}'(0) +\frac{\sw^2}{2}\Pi_{BB}'(0) \Bigr] \cwb, \\
\sw &=& \frac{g'}{\sqrt{g^2+g'^2}} = \Bigl[1 +\frac{\cw^2}{2}\Pi_{WW}'(0) -\frac{\cw^2}{2}\Pi_{BB}'(0) \Bigr] \swb, \\
Z_\mu &=& \cw W^3_\mu -\sw B_\mu = \Bigl[1 +\frac{\cw^2}{2}\Pi_{WW}'(0) +\frac{\sw^2}{2}\Pi_{BB}'(0) \Bigr] \bar Z_\mu, \\
A_\mu &=& \sw W^3_\mu +\cw B_\mu \CR
&=& \Bigl[1 +\frac{\sw^2}{2}\Pi_{WW}'(0) +\frac{\cw^2}{2}\Pi_{BB}'(0) \Bigr] \bar A_\mu +\cw\sw\Bigl[\Pi_{WW}'(0)-\Pi_{BB}'(0)\Bigr] \bar Z_\mu,
\eeqan
\eseqn
where $\bar Z_\mu = \cwb \bar W^3_\mu -\swb \bar B_\mu, \bar A_\mu = \swb \bar W^3_\mu +\cwb \bar B_\mu$. After the redefinitions, \eqref{LV2Pi} becomes
\beqa
\L_{V^2} &=& \Bigl(\frac{\bar g\bar v}{2}\Bigr)^2 \bar W^+_\mu \bar W^{-\mu} +\biggl[1 -\frac{\Pi_{WW}(0)-\Pi_{33}(0)}{m_W^2}\biggr] \frac{1}{2} \Bigl(\frac{\bar g\bar v}{2\cwb}\Bigr)^2 \bar Z_\mu \bar Z^\mu \CR
&& -\frac{1}{2}\bar G^A_\mu \hat K^{\mu\nu} \bar G^A_\nu -\bar W^+_\mu \hat K^{\mu\nu} \bar W^-_\nu \CR
&& -\bigl[1 +\Pi_{WW}'(0) -\Pi_{33}'(0)\bigr] \frac{1}{2}\bar W^3_\mu \hat K^{\mu\nu} \bar W^3_\nu +\Pi_{3B}'(0) \bar W^3_\mu \hat K^{\mu\nu} \bar B_\nu -\frac{1}{2} \bar B_\mu \hat K^{\mu\nu} \bar B_\nu \CR
&& +\frac{1}{2}\Bigl[\Pi_{GG}''(0) \frac{1}{2}\bar G^A_\mu \hat K^{2\mu\nu} \bar G^A_\nu +\Pi_{WW}''(0) \bar W^+_\mu \hat K^{2\mu\nu} \bar W^-_\nu +\Pi_{33}''(0) \frac{1}{2}\bar W^3_\mu \hat K^{2\mu\nu} \bar W^3_\nu \CR
&& \qquad +\Pi_{3B}''(0) \bar W^3_\mu \hat K^{2\mu\nu} \bar B_\nu +\Pi_{BB}''(0) \frac{1}{2}\bar B_\mu \hat K^{2\mu\nu} \bar B_\nu \Bigr],
\eeqan
where we have used $\Pi_{33} = \cw^2\Pi_{ZZ} +2\cw\sw\Pi_{Z\gamma} +\sw^2\Pi_{\gamma\gamma}$ and $\Pi_{Z\gamma}(0)=\Pi_{\gamma\gamma}(0)=0$. It is straightforward to read off the Taylor expansion coefficients of the self-energies of the redefined (barred) fields,
\bseq
\beqa
&& \bar\Pi_{WW}(0) = 0,\CR
&& \bar\Pi_{33}(0) = \cw^2\bar\Pi_{ZZ}(0) = \Pi_{33}(0)-\Pi_{WW}(0) = \cw^2\Pi_{ZZ}(0)-\Pi_{WW}(0), \\
&& \bar\Pi_{GG}'(0) = \bar\Pi_{WW}'(0) = \bar\Pi_{BB}'(0) = 0,\CR
&& \bar\Pi_{33}'(0) = \Pi_{33}'(0)-\Pi_{WW}'(0),\quad \bar\Pi_{3B}'(0) = \Pi_{3B}'(0), \\
&& \bar\Pi_{VV'}''(0) = \Pi_{VV'}''(0).
\eeqan
\eseqn
Plugging in \eqref{PiE}, we therefore obtain the oblique parameters, defined in \eqref{obldef}, in terms of the EGGM (or equivalently \BE) basis Wilson coefficients,
\beq
\hat S = g^2(E_{WB} +\frac{1}{4}E_W +\frac{1}{4}E_B), \quad
\hat T = E_T, \quad
W = \frac{g^2}{4}E_{2W}, \quad
Y = \frac{g^2}{4}E_{2B}, \quad
Z = \frac{g^2}{4}E_{2G}.
\eeq{obl}
These 5 oblique parameters constitute a subset of the 16 universal parameters.

\subsection{Triple-gauge couplings}

The field and parameter redefinitions in section~\ref{sec:utc-obl} reduce the triple gauge interactions $\L_{V^3}$ in \eqref{LV3} to the following form,
\beqa
\L_{V^3} &=& i\bar g \biggl\{ (\bar W^+_{\mu\nu}\bar W^{-\mu}-\bar W^-_{\mu\nu}\bar W^{+\mu}) \biggl[ \Bigl( 1-\frac{g^2}{4\cw^2}E_W \Bigr) \cwb \bar Z^\nu +\swb\bar A^\nu \biggr] \CR
&& +\frac{1}{2}\bar W^+_{[\mu,}\bar W^-_{\nu]} \biggl[ \Bigl(1-g'^2E_{WB} -\frac{g^2}{4\cw^2}E_W \Bigr) \cwb\bar Z^{\mu\nu} +(1 +g^2E_{WB}) \swb\bar A^{\mu\nu} \biggr] \CR
&& -\frac{E_{3W}}{v^2} \bar W^{+\nu}_\mu \bar W^{-\rho}_\nu (\cwb\bar Z_\rho ^{\,\,\,\mu}+\swb\bar A_\rho^{\,\,\,\mu}) \biggr\} +\frac{E_{2W}}{v^2}\hat K\circ \L_{\bar W\bar W\bar V}^{\SM} \CR
&& +\L_{\bar G^3}^{\SM} +\frac{E_{3G}}{v^2}\frac{g_s}{6}f^{ABC}\bar G_\mu^{A\nu}\bar G_\nu^{B\rho}\bar G_\rho^{C\mu} +\frac{E_{2G}}{v^2}\hat K\circ \L_{\bar G^3}^{\SM}.
\eeqa{LV3bar}
The terms in curly brackets correspond to the standard anomalous TGC parametrization~\cite{Hagiwara:1986vm},
\beqa
\L_{V^3} &=& i\bar g \Bigl\{ (\bar W^+_{\mu\nu}\bar W^{-\mu}-\bar W^-_{\mu\nu}\bar W^{+\mu}) \bigl[ (1+\Dgzb) \cwb \bar Z^\nu +(1+\Dgab)\swb\bar A^\nu \bigr] \CR
&& +\frac{1}{2}\bar W^+_{[\mu,}\bar W^-_{\nu]} \bigl[ (1+\Dkapzb) \cwb\bar Z^{\mu\nu} +(1 +\Dkapab) \swb\bar A^{\mu\nu} \bigr] \CR
&& +\frac{1}{m_W^2} \bar W^{+\nu}_\mu \bar W^{-\rho}_\nu (\lamzb\cwb\bar Z_\rho ^{\,\,\,\mu}+\lamab\swb\bar A_\rho^{\,\,\,\mu}) \Bigr\} +\dots
\eeqan
It is well-known that at the dimension-6 level,
\beq
\Dgab = 0,\quad \Dkapzb = \Dgzb -\frac{\sw^2}{\cw^2}\Dkapab,\quad \lamzb = \lamab,
\eeqn
which are seen to hold from \eqref{LV3bar}. The independent nonzero anomalous TGC parameters, on the other hand, can be expressed in terms of the EGGM basis Wilson coefficients as follows,
\beq
\Dgzb = -\frac{g^2}{4\cw^2}E_W, \quad
\Dkapab = g^2E_{WB}, \quad
\lamab = -\frac{g^2}{4}E_{3W}, \quad
\lamgb = -\frac{g^2}{4}E_{3G},
\eeq{tgc}
where we have defined $\lamgb$ for the triple-gluon vertex in analogy to $\lamab$. These 4 anomalous TGC parameters constitute a second subset of the universal parameters. Up to now we have introduced 9 of the 16 universal parameters.

Note that we have put bars on the anomalous TGC parameters, indicating they are defined with respect to the barred fields $\bar W^\pm_\mu, \bar Z_\mu, \bar A_\mu$. In the presence of a nonzero $\hat S$ parameter, there is kinetic mixing between $\bar W^3_\mu$ and $\bar B_\mu$, and hence between $\bar Z_\mu$ and $\bar A_\mu$ [see section~\ref{sec:utc-obl}, or \eqref{Lutc} below]. Thus, in this case the barred fields do not correspond to the physical particles, and the anomalous TGC parameters defined here are not equivalent to the usually used ones defined for the physical particles. However, the barred parameters $\Dgzb, \Dkapab, \lamab$ are more convenient for universal theories, since they can be used in parallel with the oblique parameters $\hat S, \hat T, W, Y$; see~\cite{Wells:2015eba} for a demonstration in $e^+e^-\to W^+W^-$. We will work out the relations between $\Dgzb, \Dkapab, \lamab$ and the anomalous TGC parameters in the Higgs basis, which agree with the usually adopted definitions, in section~\ref{sec:pheno} [see \eqref{LTGC} and table~\ref{tab:hb}].

\subsection{Higgs boson couplings}

The Higgs boson kinetic terms in \eqref{Lh} can be made canonical by the following field redefinition,
\beq
h = \biggl[1 -\frac{1}{2}E_H\Bigl(1+\frac{\bar h}{\bar v}+\frac{\bar h^2}{3\bar v^2}\Bigr)\biggr]\bar h.
\eeq{hredef}
We also rescale the Higgs self-coupling $\lambda$ in the SM Lagrangian,
\beq
\lambda = \Bigl(1+\frac{3}{2}E_6+E_H\Bigr)\bar\lambda,
\eeq{lamredef}
such that the tree-level relation $m_h^2=2\bar\lambda\bar v^2$ is preserved. As a result,
\beq
\L_h = \frac{1}{2}\partial_\mu \bar h \partial^\mu \bar h -\frac{1}{2}(2\bar\lambda\bar v^2) \bar h^2 -\Bigl(1-E_6-\frac{3}{2}E_H\Bigr) \bar\lambda\bar v\bar h^3 +\O(\bar h^4),
\eeqn
where we have used $v=\bar v$; see \eqref{vredef}, \eqref{PiE0}. With the redefinitions \eqref{hredef} and \eqref{lamredef}, all the new physics modifications to $\L_h$ are encoded in the momentum-independent Higgs boson self-interactions. In particular, the correction to the triple-Higgs coupling can lead to observable effects in double-Higgs production~\cite{Dolan:2012rv,Papaefstathiou:2012qe,Baglio:2012np,Goertz:2013kp,Barr:2013tda,Barger:2013jfa,Maierhofer:2013sha,deLima:2014dta,Chen:2014xra,Goertz:2014qta,Barr:2014sga,Azatov:2015oxa,Li:2015yia,Dawson:2015oha,Lu:2015jza,Dall'Osso:2015aia}. We parametrize the deviation from the SM by defining $1+\Delta\kappa_3$ to be the coefficient of $-\bar\lambda\bar v\bar h^3$, with
\beq
\Delta\kappa_3 = -E_6-\frac{3}{2}E_H
\eeq{Dk3}
in the EGGM basis. This is the 10th universal parameter.

We next consider the Higgs-fermion interactions. It is clear from \eqref{Lhf} that the tree-level relation $m_{f'}=\frac{\bar y_{f'}\bar v}{\sqrt{2}}$ is preserved if we redefine the Yukawa couplings
\beq
y_{f'} = \bar y_{f'}\Bigl(1+\frac{1}{2}E_y\Bigr).
\eeqn
Also taking into account \eqref{hredef}, we have
\beq
\L_{hf} = -\Bigl[1 +(1-E_y-\frac{1}{2}E_H)\frac{\bar h}{\bar v} -(\frac{3}{2}E_y+\frac{1}{2}E_H)\frac{\bar h^2}{\bar v^2} +\O(\bar h^3)\Bigr] \sum_{f'} \frac{\bar y_{f'} \bar v}{\sqrt{2}} \bar f' f'.
\eeqn
Defining $1+\Delta\kappa_F$ to be the coefficient of $-\frac{\bar y_{f'}}{\sqrt{2}}\bar h\bar f' f'$, we have, in the EGGM basis,
\beq
\Delta\kappa_F = -E_y-\frac{1}{2}E_H.
\eeq{DkF}
This is the 11th universal parameter.

At this point, there is no more freedom to redefine fields or parameters. In terms of the barred fields and parameters, the Higgs-vector boson interactions with zero derivatives, namely the first two terms in \eqref{LhV}, become
\beqa
\L_{hV}^{\partial^0} &=& \Bigl(1-\frac{1}{2}E_H\Bigr)\frac{2\bar h}{\bar v} \biggl[\Bigl(\frac{\bar g\bar v}{2}\Bigr)^2\bar W^+_\mu\bar W^{-\mu} +(1-2E_T)\frac{1}{2}\Bigl(\frac{\bar g\bar v}{2\cwb}\Bigr)^2\bar Z_\mu\bar Z^\mu\biggr]\CR
&& +(1-2E_H)\frac{\bar h^2}{\bar v^2} \biggl[\Bigl(\frac{\bar g\bar v}{2}\Bigr)^2\bar W^+_\mu\bar W^{-\mu} +(1-6E_T)\frac{1}{2}\Bigl(\frac{\bar g\bar v}{2\cwb}\Bigr)^2\bar Z_\mu\bar Z^\mu\biggr] \CR
&& +\O(\bar h^3\bar V^2).
\eeqan
These terms represent the rescaling of the $hVV, hhVV$ vertices in the SM. Following the common practice in the literature, we can define $1+\DkapVb$ to be the rescaling factor of the $hWW$ vertex (for the barred fields and parameters), with the 12th universal parameter
\beq
\DkapVb=-\frac{1}{2}E_H
\eeq{DkV}
in the EGGM basis. The $hZZ$ vertex is rescaled by the same factor in the absence of a nonzero $\hat T$ parameter [recall $\hat T=E_T$, see \eqref{obl}]. The two-derivative terms in \eqref{LhV}, on the other hand, represent anomalous $hVV, hhVV$ interactions with different Lorentz structures as in the SM. Since they are already of order $\frac{v^2}{\Lambda^2}$, one can directly replace the unbarred fields and parameters by the barred ones in these terms. We define parameters $f_{gg}, f_{ww}, f_{zz}, f_{z\gamma}, f_{\gamma\gamma}, f_{w\square}, f_{z\square}, f_{\gamma\square}$ corresponding to these vertices, with normalization conventions shown in \eqref{Lutc} below. Their expressions in terms of the EGGM basis Wilson coefficients can be read off from \eqref{LhV}. These parameters are not all independent of each other and of the 12 previously-defined universal parameters. In fact, they only contribute 3 more independent parameters. We will choose $f_{gg}$, $f_{z\gamma}$, $f_{\gamma\gamma}$ to be included in the universal parameters set, motivated by their close connection to the most accessible 
Higgs processes $gg\to h$, $h\to Z\gamma$, $h\to\gamma\gamma$.\footnote{These processes are the most accessible from the SMEFT point of view. In particular, though current data is not yet sensitive to $h\to Z\gamma$ at the level of SM prediction, strong constraints have been derived on the effective operators contributing to this process~\cite{Pomarol:2013zra}.} In the EGGM basis, they read,
\beqa
&& f_{gg} = 4E_{GG},\quad f_{z\gamma} = 2[2\cw^2E_{WW} -2\sw^2E_{BB} -(\cw^2-\sw^2)E_{WB}],\CR
&& f_{\gamma\gamma} = 4(E_{WW}+E_{BB}-E_{WB}).
\eeqa{fVV}
The final universal parameter is associated with the $\O(y_f^2)$ four-fermion interaction in \eqref{L4f}, and we simply define
\beq
c_{2y} = E_{2y}.
\eeq{c2y}

\subsection{Summary}
\label{sec:utc-sum}

To summarize, universal theories are characterized by the following effective Lagrangian (in the unitary gauge),
\beqa
\L_{\text{universal}} &=& \Bigl(\frac{\bar g\bar v}{2}\Bigr)^2 \bar W^+_\mu \bar W^{-\mu} +(1-\hat T)\frac{1}{2} \Bigl(\frac{\bar g\bar v}{2\cwb}\Bigr)^2 \bar Z_\mu \bar Z^\mu \CR
&& -\frac{1}{2}\bar G^A_\mu \hat K^{\mu\nu} \bar G^A_\nu -\bar W^+_\mu \hat K^{\mu\nu} \bar W^-_\nu -\frac{1}{2} \bar W^3_\mu \hat K^{\mu\nu} \bar W^3_\nu -\hat S\frac{\swb}{\cwb} \bar W^3_\mu \hat K^{\mu\nu} \bar B_\nu -\frac{1}{2} \bar B_\mu \hat K^{\mu\nu} \bar B_\nu \CR
&& -\frac{1}{m_W^2}\Bigl[Z \frac{1}{2}\bar G^A_\mu \hat K^{2\mu\nu} \bar G^A_\nu +W \Bigl(\bar W^+_\mu \hat K^{2\mu\nu} \bar W^-_\nu +\frac{1}{2}\bar W^3_\mu \hat K^{2\mu\nu} \bar W^3_\nu\Bigr) +Y \frac{1}{2}\bar B_\mu \hat K^{2\mu\nu} \bar B_\nu \Bigr] \CR
&& +i\bar g \Bigl\{ (\bar W^+_{\mu\nu}\bar W^{-\mu}-\bar W^-_{\mu\nu}\bar W^{+\mu}) \bigl[ (1+\Dgzb) \cwb \bar Z^\nu +\swb\bar A^\nu \bigr] \CR
&&\qquad +\frac{1}{2}\bar W^+_{[\mu,}\bar W^-_{\nu]} \bigl[ (1+\Dkapzb) \cwb\bar Z^{\mu\nu} +(1 +\Dkapab) \swb\bar A^{\mu\nu} \bigr] \CR
&&\qquad +\frac{\lamab}{m_W^2} \bar W^{+\nu}_\mu \bar W^{-\rho}_\nu (\cwb\bar Z_\rho ^{\,\,\,\mu}+\swb\bar A_\rho^{\,\,\,\mu}) \Bigr\} +\frac{W}{m_W^2}\hat K\circ \L_{\bar W\bar W\bar V}^{\SM} \CR
&&\qquad +\L_{\bar G^3}^{\SM} -\frac{\lamgb}{m_W^2}\frac{\bar g_s}{6}f^{ABC}\bar G_\mu^{A\nu}\bar G_\nu^{B\rho}\bar G_\rho^{C\mu} +\frac{Z}{m_W^2}\hat K\circ \L_{\bar G^3}^{\SM} \CR
&& +\frac{1}{2}\partial_\mu \bar h \partial^\mu \bar h -\frac{1}{2}(2\bar\lambda\bar v^2) \bar h^2 -(1+\Delta\kappa_3) \bar\lambda\bar v\bar h^3 \CR
&& -\Bigl[1 +(1+\DkapFb)\frac{\bar h}{\bar v} +\Bigl(\frac{3}{2}\DkapFb-\frac{1}{2}\DkapVb\Bigr)\frac{\bar h^2}{\bar v^2}\Bigr] \sum_{f'} \frac{\bar y_{f'} \bar v}{\sqrt{2}} \bar f' f' \CR
&& +(1+\DkapVb)\frac{2\bar h}{\bar v} \biggl[\Bigl(\frac{\bar g\bar v}{2}\Bigr)^2\bar W^+_\mu\bar W^{-\mu} +(1-2\hat T)\frac{1}{2}\Bigl(\frac{\bar g\bar v}{2\cwb}\Bigr)^2\bar Z_\mu\bar Z^\mu\biggr]\CR
&&\qquad +(1+4\DkapVb)\frac{\bar h^2}{\bar v^2} \biggl[\Bigl(\frac{\bar g\bar v}{2}\Bigr)^2\bar W^+_\mu\bar W^{-\mu} +(1-6\hat T)\frac{1}{2}\Bigl(\frac{\bar g\bar v}{2\cwb}\Bigr)^2\bar Z_\mu\bar Z^\mu\biggr] \CR
&& +\Bigl(\frac{\bar h}{\bar v}+\frac{\bar h^2}{2\bar v^2}\Bigr) \Bigl[ f_{gg}\frac{\bar g_s^2}{4} \partial_{[\mu,}\bar G^A_{\nu]} \partial^{[\mu,}\bar G^{A\nu]} +f_{ww}\frac{\bar g^2}{2} \bar W^+_{\mu\nu}\bar W^{-\mu\nu} +f_{zz}\frac{\bar g^2}{4\cwb^2}\bar Z_{\mu\nu}\bar Z^{\mu\nu} \CR
&&\qquad +f_{z\gamma}\frac{\bar g\bar g'}{2}\bar Z_{\mu\nu}\bar A^{\mu\nu} +f_{\gamma\gamma}\frac{\bar e^2}{4} \bar A_{\mu\nu}\bar A^{\mu\nu} +f_{w\square}\bar g^2(\bar W^-_\mu\partial_\nu \bar W^{+\mu\nu}+\text{h.c.}) \CR
&&\qquad +f_{z\square}\bar g^2\bar Z_\mu\partial_\nu\bar Z^{\mu\nu} +f_{\gamma\square}\bar g\bar g'\bar Z_\mu\partial_\nu\bar A^{\mu\nu}\Bigr] \CR
&& +c_{2y} J_{y\alpha}^\dagger J_y^\alpha +\sum_{f} i\bar f\gamma^\mu D_\mu f +\O(\bar V^4, \bar h^4, \bar h^3 f^2, \bar h^3\bar V^2, \bar h\bar V^3),
\eeqa{Lutc}
where $\hat K^{\mu\nu}$, $\hat K^{2\mu\nu}$ are defined in \eqref{Kdef}, and the action of $\hat K\circ$ is shown in \eqref{Kcirc} and the discussions below that equation. $f'$ denotes mass eigenstates, while $f\in\{q,l,u,d,e\}$ denotes gauge eigenstates. They agree with each other except for $d_L$ in the $SU(2)_L$ doublet $q$, for which $d_L=\VCKM d'_L$. The scalar current $J_y^\alpha$ is defined in \eqref{Jydef}. The gauge interactions of $f$ from $i\bar f\gamma^\mu D_\mu f$ are the same as in the SM, shown in \eqref{LVffSM}, with unbarred fields and parameters replaced by barred ones. 

Corresponding to the 16 independent Wilson coefficients in each basis, we have defined 16 parameters that conveniently characterize all the indirect effects of universal theories, dubbed universal parameters. They include:
\begin{itemize}
\item 5 oblique parameters $\hat S$, $\hat T$, $W$, $Y$, $Z$;
\item 4 anomalous TGC parameters $\Dgzb$, $\Dkapab$, $\lamab$, $\lamgb$;
\item 3 parameters for the rescaling of the SM $h^3$, $hff$, $hVV$ couplings $\Delta\kappa_3$, $\DkapFb$, $\DkapVb$;
\item 3 parameters for the $hVV$ couplings with non-SM Lorentz structures $f_{gg}$, $f_{z\gamma}$, $f_{\gamma\gamma}$;
\item 1 parameter for the $\O(y_f^2)$ four-fermion coupling $c_{2y}$.
\end{itemize}
Eq.~\eqref{Lutc} can be viewed as the definition of these parameters: they are defined from the terms in the effective Lagrangian when $\L_{\text{universal}}$ is cast in the form shown in this equation by field and parameter redefinitions.

Each of the 16 universal parameters can be expressed as a linear combination of Wilson coefficients in a particular SMEFT basis (in a sense they constitute an alternative basis for universal theories). We have shown in detail how to derive the expressions in the EGGM basis. The results are presented in eqs.~\eqref{obl}, \eqref{tgc}, \eqref{Dk3}, \eqref{DkF}, \eqref{DkV}, \eqref{fVV}, \eqref{c2y}, and summarized in the second column of table~\ref{tab:utc}. Applying the basis transformation formulas tabulated in section~\ref{sec:def}, we arrive at the following columns of table~\ref{tab:utc}, showing how the universal parameters should be written down in each basis. In particular, we note that in the SILH and Warsaw bases, Wilson coefficients of fermionic operators enter the oblique parameters when the latter are defined according to the procedure described at the beginning of section~\ref{sec:utc-obl}. In fact, they correspond to combinations of fermionic operators allowed in universal theories whose effects on observables are equivalent to vector boson self-energy corrections. To consistently use the constraints on the oblique parameters, the fermionic operators should be traded for their bosonic counterparts, and their contributions to the oblique parameters evaluated. 

\begin{sidewaystable}[tbp]
\begin{small}
\begin{tabular}{|l|l|l|l|l|l|}
\hline
 & EGGM & SILH & Warsaw & \BE & \BS \\
\hline
$\hat S$ & $g^2(E_{WB}$ & $g^2(\frac{1}{4}S_W +\frac{1}{4}S_B$ & $g^2(\frac{1}{gg'}C_{HWB} +\frac{1}{4}C_{HJW} +\frac{1}{4}C_{HJB}$ &  $g^2(\bar E_{WB}$ & $g^2(\frac{1}{4}\bar S_W +\frac{1}{4}\bar S_B)$  \\
 & $\quad +\frac{1}{4}E_W +\frac{1}{4}E_B)$  & $\quad -\frac{1}{2}S_{2JW} -\frac{1}{2}S_{2JB})$ & $\quad -\frac{1}{2}C_{2JW} -\frac{1}{2}C_{2JB})$ & $\quad +\frac{1}{4}\bar E_W +\frac{1}{4}\bar E_B)$ & \\
$\hat T$ & $E_T$ & $S_T -\frac{g'^2}{2}S_{2JB}$ & $-\frac{1}{2}C_{HD} +\frac{g'^2}{2}(C_{HJB}-C_{2JB})$ &  $\bar E_T$ & $\bar S_T$ \\
$W$ & $\frac{g^2}{4}E_{2W}$ & $-\frac{g^2}{2}S_{2JW}$ & $-\frac{g^2}{2}C_{2JW}$ & $\frac{g^2}{4}\bar E_{2W}$ & $\frac{g^2}{4}\bar S_{2W}$ \\
$Y$ & $\frac{g^2}{4}E_{2B}$ & $-\frac{g^2}{2}S_{2JB}$ & $-\frac{g^2}{2}C_{2JB}$ & $\frac{g^2}{4}\bar E_{2B}$ & $\frac{g^2}{4}\bar S_{2B}$ \\
$Z$ & $\frac{g^2}{4}E_{2G}$ & $-\frac{g^2}{2}S_{2JG}$ & $-\frac{g^2}{2}C_{2JG}$ & $\frac{g^2}{4}\bar E_{2G}$ & $\frac{g^2}{4}\bar S_{2G}$ \\
\hline
$\Dgzb$ & $-\frac{g^2}{4\cw^2}E_W$ & $-\frac{g^2}{4\cw^2}(S_W+S_{HW}$ & $-\frac{g^2}{4\cw^2}(C_{HJW}-2C_{2JW})$ & $-\frac{g^2}{4\cw^2}\bar E_W$ & $-\frac{g^2}{4\cw^2}(\bar S_W+\bar S_{HW})$ \\
 & & $\quad -2S_{2JW})$ & & & \\
$\Dkapab$ & $g^2E_{WB}$ & $-\frac{g^2}{4}(S_{HW}+S_{HB})$ & $\frac{\cw}{\sw}C_{HWB}$ & $g^2\bar E_{WB}$ & $-\frac{g^2}{4}(\bar S_{HW}+\bar S_{HB})$ \\
$\lamab$ & $-\frac{g^2}{4}E_{3W}$ & $-\frac{g^2}{4}S_{3W}$ & $-\frac{3g}{2}C_W$ & $-\frac{g^2}{4}\bar E_{3W}$ & $-\frac{g^2}{4}\bar S_{3W}$ \\
$\lamgb$ & $-\frac{g^2}{4}E_{3G}$ & $-\frac{g^2}{4}S_{3G}$ & $-\frac{3g^2}{2g_s}C_G$ & $-\frac{g^2}{4}\bar E_{3G}$ & $-\frac{g^2}{4}\bar S_{3G}$ \\
\hline
$\Delta\kappa_3$ & $-E_6 -\frac{3}{2}E_H$ & $-S_6 -\frac{3}{2}S_H +\frac{g^2}{4}S_{2JW}$ & $-\frac{1}{\lambda}C_H +3C_{H\square} -\frac{3}{4}C_{HD}$ & $-\bar E_6 -\frac{3}{2}\bar E_H$ & $-\bar S_6 -\frac{3}{2}\bar S_H$ \\
 &  &  & $\quad -\frac{g^2}{4}(C_{HJW}-C_{2JW})$ & $\quad -\frac{1}{2}\bar E_r -4\lambda\bar E_{K4}$ & $\quad -\frac{1}{2}\bar S_r -4\lambda\bar S_{K4}$ \\
$\DkapFb$ & $-E_y -\frac{1}{2}E_H$ & $-S_y -\frac{1}{2}S_H +\frac{g^2}{4}S_{2JW}$ & $-C_y +C_{H\square} -\frac{1}{4}C_{HD}$ & $-\frac{1}{2}\bar E_H -2\lambda\bar E_{K4}$ & $-\frac{1}{2}\bar S_H -2\lambda\bar S_{K4}$ \\
 & & & $\quad -\frac{g^2}{4}(C_{HJW}-C_{2JW})$ &  &  \\
$\DkapVb$ & $-\frac{1}{2}E_H$ & $-\frac{1}{2}S_H +\frac{3g^2}{4}S_{2JW}$ & $C_{H\square} -\frac{1}{4}C_{HD}$ & $-\frac{1}{2}(\bar E_H-\bar E_r)$ & $-\frac{1}{2}(\bar S_H-\bar S_r)$ \\
 & & & $\quad -\frac{3g^2}{4}(C_{HJW}-C_{2JW})$ & & \\
\hline
$f_{gg}$ & $4E_{GG}$ & $4S_{GG}$ & $\frac{4}{g_s^2}C_{HG}$ &  $4\bar E_{GG}$ & $4\bar S_{GG}$  \\
$f_{z\gamma}$ & $2[2\cw^2E_{WW} -2\sw^2E_{BB}$ & $-4\sw^2S_{BB}$ & $\frac{2}{gg'}[2\cw\sw(C_{HW}-C_{HB})$ &  $2[2\cw^2\bar E_{WW} -2\sw^2\bar E_{BB}$ & $-4\sw^2\bar S_{BB}$  \\
 & $\quad -(\cw^2-\sw^2)E_{WB}]$ & $\quad -\frac{1}{2}(S_{HW}-S_{HB})$ & $\quad -(\cw^2-\sw^2)C_{HWB}]$ & $\quad -(\cw^2-\sw^2)\bar E_{WB}]$ & $\quad -\frac{1}{2}(\bar S_{HW}-\bar S_{HB})$ \\
$f_{\gamma\gamma}$ & $4(E_{WW}+E_{BB}-E_{WB})$ & $4S_{BB}$ & $4(\frac{1}{g^2}C_{HW} +\frac{1}{g'^2}C_{HB} -\frac{1}{gg'}C_{HWB})$ &  $4(\bar E_{WW}+\bar E_{BB}-\bar E_{WB})$ & $4\bar S_{BB}$  \\
\hline
$c_{2y}$ & $E_{2y}$ & $S_{2y}$ & $C_{2y}$ & $\bar E_{K4}$ & $\bar S_{K4}$ \\
\hline
\end{tabular}
\end{small}
\caption{\label{tab:utc} The 16 universal parameters, defined in \eqref{Lutc}, in terms of the independent Wilson coefficients when $\L_{\text{universal}}$ is written in the EGGM, SILH, Warsaw, \BE\ and \BS\ bases as in \eqref{LuE}, \eqref{LuS}, \eqref{LuW}, \eqref{LuB}. These parameters generalize the oblique parameters framework, and constitute a complete description of universal theories.}
\end{sidewaystable}
\clearpage

The other parameters appearing in \eqref{Lutc} are related to the independent universal parameters as follows,
\bseq
\beqa
\Dkapzb &=& \Dgzb -\frac{\sw^2}{\cw^2}\Dkapab, \\
f_{ww} &=& f_{z\gamma} +\sw^2f_{\gamma\gamma} +\frac{2}{g^2}\Dkapab, \\
f_{zz} &=& (\cw^2-\sw^2)f_{z\gamma} +\cw^2\sw^2f_{\gamma\gamma} +\frac{2}{g^2}\Dkapab, \\
f_{w\square} &=& -\frac{2\cw^2}{g^2}\Dgzb, \\
f_{z\square} &=& -\frac{2}{g^2}\Bigl[(\cw^2-\sw^2)\Dgzb +\frac{\sw^2}{\cw^2}(\Dkapab-\hat S)\Bigr], \\
f_{\gamma\square} &=& -\frac{2}{g^2}(2\cw^2\Dgzb-\Dkapab+\hat S).
\eeqan
\eseq{deppar}
Also, note that the $hhff$ and $hhVV$ couplings are completely determined by the $hff$ and $hVV$ couplings, as is clear from \eqref{Lutc}. This is a consequence of the $h$ being part of the $SU(2)_L$ doublet $H$, and also holds in general nonuniversal theories.

\section{Connection to the Higgs basis}
\label{sec:pheno}

It has been recently proposed that a common SMEFT basis that is most straightforwardly connected to observables be adopted by the precision analyses community~\cite{HiggsBasis}. This proposal is motivated by the earlier idea of BSM primaries~\cite{Gupta:2014rxa}, and features a set of {\it effective couplings} that capture corrections to all the interaction vertices in the SM Lagrangian, when the following 3 {\it Higgs basis defining conditions} (not to be confused with the oblique parameters defining conditions listed in section~\ref{sec:utc-obl}) are satisfied:\footnote{The third Higgs basis defining condition is not explicitly stated in a complete way in the current version of~\cite{HiggsBasis}, where the prescription for the $h^2Vff$ terms is not specified. But it is clear from the calculations in~\cite{HiggsBasis} that the condition stated here is implicitly assumed.}
\begin{itemize}
\item[1)]
  All the mass eigenstates have canonically normalized kinetic terms with no kinetic mixing or higher-derivative self-interactions.
\item[2)]
  The input observables $m_Z, m_H, G_F, \alpha, \alpha_s, m_f$ are not modified at leading order.
\item[3)]
  The combinations of anomalous $Vff$, $hVff$, $h^2Vff$ interactions are proportional to $(1+\frac{h}{v})^2$.
\end{itemize}
One can choose a subset of these effective couplings to be independent couplings, and the rest are dependent couplings due to the correlations of new physics effects at the dimension-6 level with linearly-realized electroweak symmetry breaking. The set of independent couplings constitute a complete basis, called the Higgs basis, since they can be written as independent linear combinations of Wilson coefficients in any other basis. With a slight abuse of terminology, in the following we will refer to the ``effective couplings in the Lagrangian when the Higgs basis defining conditions are satisfied'' as ``Higgs basis couplings.'' To avoid confusion with the ``independent couplings constituting the Higgs basis'', we will call the latter simply ``independent couplings.''

Though the Higgs basis is still work in progress, and especially it is yet to be understood how to extend the framework beyond leading order, the virtue of the proposal is clear, at least at leading order. Due to the Higgs basis defining conditions specified above, all BSM effects are captured by {\it vertex corrections} involving the {\it physical particles}, and all new physics contributions to precision observables are direct (there is no indirect contribution from shifting the input observables, see~\cite{Wells:2014pga}). As a result, there is almost a one-to-one mapping between the effective couplings and many precision observables. 

While the Higgs basis proposal is largely motivated by a convenient characterization of indirect BSM effects in generic nonuniversal theories, it is helpful to work out the Higgs basis couplings in the special case of universal theories, as we will do in section~\ref{sec:hb-c}. In this case, all the Higgs basis couplings are determined by the 16 universal parameters. This number is much smaller than the number of independent couplings in general nonuniversal theories, which means that in addition to the generally-valid coupling relations listed in~\cite{HiggsBasis} (expressions of dependent couplings in terms of independent couplings), universal theories predict relations among the {\it independent} couplings. As we will discuss in section~\ref{sec:hb-cor}, on the one hand, these relations serve as a definition of universal theories in the Higgs basis; on the other hand, the pattern of deviations from the SM predictions for the precision observables can be inferred from these correlations, which will make it clear in what sense the BSM effects are ``universal'' in universal theories.

\subsection{Higgs basis couplings in universal theories}
\label{sec:hb-c}

We will start from the Lagrangian \eqref{Lutc}, where the BSM effects are captured by the 16 universal parameters, and make further field and parameter redefinitions to satisfy the Higgs basis defining conditions. An alternative strategy is to start from the SMEFT Lagrangian in a basis that does not contain $\O_{2B}, \O_{2W}, \O_{2G}$ (and hence no higher-derivative gauge boson self-interactions) such as the SILH or Warsaw basis, namely from \eqref{LuS} or \eqref{LuW}, and follow the steps in~\cite{HiggsBasis} to redefine the fields and parameters. The resulting Higgs basis parameters can then be recast in terms of the universal parameters with the help of table~\ref{tab:utc}. We have explicitly checked that both approaches yield identical final results. In the following we will illustrate in detail the first approach, which involves the universal parameters more directly. The distinction between independent vs.\ dependent couplings is not relevant for this calculation, so we will not specify which couplings are to be chosen as independent couplings till the end of this subsection.

First, according to the first Higgs basis defining condition, the terms proportional to $W, Y, Z$ should be eliminated, since they represent higher-derivative gauge boson self-interactions. Recall from table~\ref{tab:utc} that $W,Y,Z$ are proportional to $E_{2W}, E_{2B}, E_{2G}$, respectively, so the terms to be eliminated are actually
\beq
\frac{1}{v^2}(E_{2W}\O_{2W} +E_{2B}\O_{2B} +E_{2G}\O_{2G}) = \frac{1}{m_W^2} (W\O_{2W} +Y\O_{2B} +Z\O_{2G}).
\eeqn
By \eqref{O2V}, this is equivalent to
\beq
-\frac{1}{m_W^2}(W\O_W+Y\O_B) +\frac{W}{v^2}(4\O_6+\O_y-3\O_H) -\frac{Y}{v^2}\frac{\sw^2}{\cw^2}\O_T -\frac{1}{2m_W^2}(W\O_{2JW}+Y\O_{2JB}+Z\O_{2JG}).
\eeqn
It can be directly read off from table~\ref{tab:utc} how the coefficients of $\O_W, \O_B, \O_6, \O_y, \O_H, \O_T$ contribute to the universal parameters. Thus, $\L_{\text{universal}}$ is equivalent to \eqref{Lutc} with the following replacements
\beqa
&& \hat S \to \hat S-W-Y =\De_3,\quad \hat T\to \hat T-\frac{\sw^2}{\cw^2}Y =\De_1-\De_2,\quad W, Y, Z \to 0, \CR
&& \Dgzb\to \Dgzb +\frac{W}{\cw^2} =\Dgzb -\frac{\De_2}{\cw^2},\quad \Delta\kappa_3\to \Delta\kappa_3+\frac{W}{2} =\Delta\kappa_3-\frac{\De_2}{2},\CR
&& \DkapFb\to \DkapFb+\frac{W}{2} =\DkapFb-\frac{\De_2}{2},\quad \DkapVb\to \DkapVb+\frac{3W}{2} =\DkapVb-\frac{3\De_2}{2},
\eeqa{replace}
along with the addition of the terms
\beq
-\frac{1}{2m_W^2}(W\O_{2JW}+Y\O_{2JB}+Z\O_{2JG}).
\eeq{add4f}
In \eqref{replace} we have used the parameters $\De_{1,2,3}$, defined by
\beq
\De_1\equiv \hat T-W-\frac{\sw^2}{\cw^2}Y,\quad \De_2\equiv -W,\quad \De_3\equiv \hat S-W-Y.
\eeq{Dedef}
These are the three independent linear combinations of $\hat S, \hat T, W, Y$ that enter the pole observables, which have been used historically~\cite{Altarelli:1990zd,Altarelli:1991fk,Barbieri:2004qk}.\footnote{As a historical note, $\De_{1,2,3}$ used to be associated with $\hat S$, $\hat T$, $\hat U$. But as argued in~\cite{Barbieri:2004qk}, $\hat U$ is generically higher order compared with $W$ and $Y$ if there is a separation of scales $\Lambda\gg v$. Recasting the oblique parameters analyses in the SMEFT language as in~\cite{Barbieri:2004qk}, and more systematically in this paper, makes it clear that $\De_{1,2,3}$ are actually associated with linear combinations of $\hat S$, $\hat T$, $W$, $Y$ at the dimension-6 level.}

Next, we focus on the electroweak sector. The neutral vector boson kinetic terms
\beqa
&& -\frac{1}{2} \bar W^3_\mu \hat K^{\mu\nu} \bar W^3_\nu -\De_3\frac{\swb}{\cwb} \bar W^3_\mu \hat K^{\mu\nu} \bar B_\nu -\frac{1}{2} \bar B_\mu \hat K^{\mu\nu} \bar B_\nu \CR
&=& -(1-2\swb^2\De_3)\frac{1}{2}\bar Z_\mu \hat K^{\mu\nu} \bar Z_\nu -(1+2\swb^2\De_3)\frac{1}{2}\bar A_\mu \hat K^{\mu\nu} \bar A_\nu \CR
&& -\frac{\swb}{\cwb}(\cwb^2-\swb^2)\De_3\bar Z_\mu \hat K^{\mu\nu} \bar A_\nu
\eeqa{nvbkin}
can be diagonalized and canonically normalized by redefining the fields
\bseq
\beqa
\bar Z_\mu &=& (1+\swb^2\De_3)\hat Z_\mu, \\
\bar A_\mu &=& (1-\swb^2\De_3)\hat A_\mu -\frac{\swb}{\cwb}(\cwb^2-\swb^2)\De_3\hat Z_\mu.
\eeqan
\eseq{ZAredef}
Eq.~\eqref{nvbkin} then becomes $-\frac{1}{2}\hat Z_\mu \hat K^{\mu\nu} \hat Z_\nu -\frac{1}{2}\hat A_\mu \hat K^{\mu\nu} \hat A_\nu$. The $W^\pm$ fields need not be redefined, and we write $\bar W^\pm_\mu=\hat W^\pm_\mu$ so that the properly-defined fields satisfying the Higgs basis defining conditions are denoted with hats. Further, to preserve the leading-order relations between the input observables $m_Z, G_F, \alpha$ and the SM Lagrangian parameters as required by the second Higgs basis defining condition, the following parameter redefinitions are needed,
\bseq
\beqa
&& \left(\frac{\bar e\bar v}{\cwb\swb}\right)^2 = (1+\De_1-\De_2-2\sw^2\De_3)\left(\frac{\hat e\hat v}{\cwh\swh}\right)^2, \\
&& \bar v^2 = (1-\De_2)\hat v^2, \label{vredef-hb} \\
&& \bar e = (1+\sw^2\De_3)\hat e.
\eeqan
\eseq{Oinredef}
Accordingly, we have
\bseq
\beqa
&& \cwb = \biggl[1 +\frac{\sw^2}{\cw^2-\sw^2}\Bigl(\frac{\De_1}{2} -2\sw^2\De_3\Bigr)\biggr] \cwh, \\
&& \swb = \biggl[1 -\frac{\cw^2}{\cw^2-\sw^2}\Bigl(\frac{\De_1}{2} -2\sw^2\De_3\Bigr)\biggr] \swh, \\
&& \bar g = \frac{\bar e}{\swb} = \Bigl(1 +\frac{\cw^2}{\cw^2-\sw^2}\frac{\De_1}{2} -\frac{\sw^2}{\cw^2-\sw^2}\De_3\Bigr)\hat g, \\
&& \bar g' = \frac{\bar e}{\cwb} = \Bigl(1 -\frac{\sw^2}{\cw^2-\sw^2}\frac{\De_1}{2} +\frac{\sw^2}{\cw^2-\sw^2}\De_3\Bigr)\hat g',
\eeqan
\eseq{Oinredef2}
where the first two equations follow from \eqref{Oinredef} and $\cwh^2+\swh^2=1$. These parameter redefinitions ensure that
\bseq
\beqa
\Delta\L_{m_Z} &=& (1-\De_1+\De_2)\frac{1}{2} \Bigl(\frac{\bar e\bar v}{2\cwb\swb}\Bigr)^2 \bar Z_\mu \bar Z^\mu = \frac{1}{2} \Bigl(\frac{\hat e\hat v}{2\cwh\swh}\Bigr)^2 \hat Z_\mu \hat Z^\mu \CR
&&\Rightarrow m_Z^{\text{LO}} = \frac{\hat e\hat v}{2\cwh\swh}, \\
\Delta\L_{G_F} &=& \Bigl(\frac{\bar g\bar v}{2}\Bigr)^2 \bar W^+_\mu \bar W^{-\mu} +\frac{\bar g}{\sqrt{2}}(\bar W^+_\mu\bar l_i\gamma^\mu\sigma^+l_i +\text{h.c.}) +\frac{\De_2}{v^2}(\bar l_i\gamma_\mu l_j)(\bar l_j\gamma^\mu l_i) \CR
&&\Rightarrow -2\sqrt{2}G_F^{\text{LO}} = -\frac{2}{\bar v^2} +\frac{2\De_2}{\bar v^2} = -\frac{2}{\hat v^2},\label{GFLO} \\
\Delta\L_{\alpha} &=& \bar e\bar A_\mu \sum_f Q_f \bar f \gamma^\mu f \supset \hat e\hat A_\mu \sum_f Q_f \bar f \gamma^\mu f \CR
&&\Rightarrow \alpha^{\text{LO}} = \frac{\hat e^2}{4\pi}.
\eeqan
\eseqn
In deriving \eqref{GFLO}, we have noticed that $-2\sqrt{2}G_F^{\text{LO}}$ is identified as the coefficient of the effective four-fermion interaction term $(\bar e_L\gamma_\rho\nu_e)(\bar\nu_\mu\gamma^\rho\mu_L) +\text{h.c.}$ after the $W^\pm$ propagator is integrated out. The first two terms in $\Delta\L_{G_F}$ are the same as the corresponding SM terms with barred fields and parameters [we have defined $\sigma^+=(\sigma^1+i\sigma^2)/2$], which contribute $-\frac{2}{\bar v^2}$ to this coefficient, while the third term contains
\beq
\frac{2\De_2}{v^2}\bigl[(\bar e_L\gamma_\rho\mu_L)(\bar\nu_\mu\gamma^\rho\nu_e) +\text{h.c.}\bigr] = \frac{2\De_2}{v^2}\bigl[(\bar e_L\gamma_\rho\nu_e)(\bar\nu_\mu\gamma^\rho\mu_L) +\text{h.c.}\bigr],
\eeqn
where a Fierz rearrangement has been made.

As a consequence of the field and parameter redefinitions above, the $W$ boson mass term becomes
\beq
\Bigl(\frac{\bar g\bar v}{2}\Bigr)^2 \bar W^+_\mu \bar W^{-\mu} = \Bigl[(1+\delta_m)\frac{\hat g\hat v}{2}\Bigr]^2 \hat W^+_\mu \hat W^{-\mu},
\eeqn
where
\beq
\delta_m = \frac{\cw^2}{\cw^2-\sw^2}\frac{\De_1}{2} -\frac{\De_2}{2} -\frac{\sw^2}{\cw^2-\sw^2}\De_3
\eeqn
is one of the the Higgs basis couplings.\footnote{This parameter is denoted by $\delta m$ in the current version of~\cite{HiggsBasis}. We prefer $\delta_m$ because $\delta m$ is often used to refer to the absolute shift, rather than the factional shift, of a mass.} Also, using \eqref{ZAredef}, \eqref{Oinredef}, \eqref{Oinredef2}, we obtain the charged-current (CC) and neutral-current (NC) interactions of the SM fermions, and the triple-gauge interactions,
\bseq
\beqa
\L_{\text{CC}} &=& \frac{\bar g}{\sqrt{2}}\bigl[\bar W^+_\mu(\bar q_i\gamma^\mu\sigma^+q_i +\bar l_i\gamma^\mu\sigma^+l_i) +\text{h.c.}\bigr] \CR
&=& \Bigl(1+\frac{\cw^2}{\cw^2-\sw^2}\frac{\De_1}{2} -\frac{\sw^2}{\cw^2-\sw^2}\De_3\Bigr)\frac{\hat g}{\sqrt{2}}\bigl[\hat W^+_\mu(\bar q_i\gamma^\mu\sigma^+q_i +\bar l_i\gamma^\mu\sigma^+l_i) +\text{h.c.}\bigr]\CR
&\equiv& \frac{\hat g}{\sqrt{2}}\Bigl\{\hat W^+_\mu\Bigl[\bigl(1+[\dgL^{Wq}]_{ij}\bigr)\bar q_i\gamma^\mu\sigma^+q_j +\bigl(1+[\dgL^{Wl}]_{ij}\bigr)\bar l_i\gamma^\mu\sigma^+l_j\Bigr] +\text{h.c.}\Bigr\},\label{LCC} \\
\L_{\text{NC}} &=& \sum_f \Bigl[\frac{\bar e}{\cwb\swb}\bar Z_\mu(T^3_f- Q_f\swb^2) +\bar e\bar A_\mu Q_f \Bigr] \bar f_i\gamma^\mu f_i \CR
&=& \sum_f \biggl\{\frac{\hat e}{\cwh\swh}\hat Z_\mu\biggl[T^3_f\Bigl(1+\frac{\De_1}{2}\Bigr) - Q_f\swh^2\Bigl(1-\frac{1}{\cw^2-\sw^2}\bigl(\frac{\De_1}{2}-\De_3\bigr)\Bigr)\biggr] +\hat e\hat A_\mu Q_f \biggr\} \bar f_i\gamma^\mu f_i \CR
&\equiv& \sum_f \Bigl[\frac{\hat e}{\cwh\swh}\hat Z_\mu\bigl(T^3_f-Q_f\swh^2 +[\delta g_{L/R}^{Zf}]_{ij}\bigr) +\hat e\hat A_\mu Q_f\delta_{ij} \Bigr] \bar f_i\gamma^\mu f_j,\label{LNC} \\
\L_{\text{TGC}} &=& i\bar g \Bigl\{ (\bar W^+_{\mu\nu}\bar W^{-\mu}-\bar W^-_{\mu\nu}\bar W^{+\mu}) \biggl[ \Bigl(1+\Dgzb-\frac{\De_2}{\cw^2}\Bigr) \cwb \bar Z^\nu +\swb\bar A^\nu \biggr] \CR
&&\,\, +\frac{1}{2}\bar W^+_{[\mu,}\bar W^-_{\nu]} \biggl[ (1+\Dkapzb) \cwb\bar Z^{\mu\nu} +(1 +\Dkapab) \swb\bar A^{\mu\nu} \biggr] \CR
&&\,\, +\frac{\lamab}{m_W^2} \bar W^{+\nu}_\mu \bar W^{-\rho}_\nu (\cwb\bar Z_\rho ^{\,\,\,\mu}+\swb\bar A_\rho^{\,\,\,\mu}) \Bigr\} +\L_{\bar G^3}^\SM -\frac{\lamgb}{m_W^2}\frac{\bar g_s}{6}f^{ABC}\bar G_\mu^{A\nu}\bar G_\nu^{B\rho}\bar G_\rho^{C\mu} \CR
&=&  i\hat g \Bigl\{ (\hat W^+_{\mu\nu}\hat W^{-\mu}-\hat W^-_{\mu\nu}\hat W^{+\mu}) \CR
&&\quad \biggl[ \Bigl(1+\Dgzb -\frac{\De_2}{\cw^2} +\frac{\sw^2}{\cw^2-\sw^2}\bigl(\frac{\De_1}{2\sw^2}-\frac{\De_3}{\cw^2}\bigr)\Bigr) \cwh \hat Z^\nu +\swh\hat A^\nu \biggr] \CR
&&\,\, +\frac{1}{2}\hat W^+_{[\mu,}\hat W^-_{\nu]} \CR
&&\quad \biggl[ \Bigl(1+\Dgzb -\frac{\sw^2}{\cw^2}\Dkapab -\frac{\De_2}{\cw^2} +\frac{\sw^2}{\cw^2-\sw^2}\bigl(\frac{\De_1}{2\sw^2}-\frac{\De_3}{\cw^2}\bigr)\Bigr) \cwh \hat Z^{\mu\nu} +(1 +\Dkapab)\swh\hat A^{\mu\nu} \biggr] \CR
&&\,\, +\frac{\lamab}{m_W^2} \hat W^{+\nu}_\mu \hat W^{-\rho}_\nu (\cwh\hat Z_\rho ^{\,\,\,\mu}+\swh\hat A_\rho^{\,\,\,\mu}) \Bigr\} +\L_{\hat G^3}^\SM -\frac{\lamgb}{m_W^2}\frac{\hat g_s}{6}f^{ABC}\hat G_\mu^{A\nu}\hat G_\nu^{B\rho}\hat G_\rho^{C\mu} \CR
&\equiv& i\hat g \Bigl\{ (\hat W^+_{\mu\nu}\hat W^{-\mu}-\hat W^-_{\mu\nu}\hat W^{+\mu}) \bigl[ (1+\dgz) \cwh \hat Z^\nu +\swh\hat A^\nu \bigr] \CR
&&\,\, +\frac{1}{2}\hat W^+_{[\mu,}\hat W^-_{\nu]} \bigl[ (1+\dkapz) \cwh\hat Z^{\mu\nu} +(1 +\dkapa) \swh\hat A^{\mu\nu} \bigr] \CR
&&\,\, +\frac{\lambda_\gamma}{m_W^2} \hat W^{+\nu}_\mu \hat W^{-\rho}_\nu (\cwh\hat Z_\rho ^{\,\,\,\mu}+\swh\hat A_\rho^{\,\,\,\mu}) \Bigr\} +\L_{\hat G^3}^\SM +\frac{c_{3G}}{v^2}\hat g_s^3f^{ABC}\hat G_\mu^{A\nu}\hat G_\nu^{B\rho}\hat G_\rho^{C\mu},\label{LTGC}
\eeqan
\eseqn
where $\dgL^{Zf}$ and $\dgR^{Zf}$ apply for $f\in\{u_L,d_L,e_L,\nu\}$ and $f\in\{u_R,d_R,e_R\}$, respectively. Note that $T^3_f=0$ for $f\in\{u_R,d_R,e_R\}$. We have also included the triple-gluon interactions in $\L_{\text{TGC}}$, with $\bar G^A_\mu=\hat G^A_\mu$, $\bar g_s=\hat g_s$. The results for the Higgs basis couplings $\dgL^{Wf}$, $\dgL^{Zf}$, $\dgR^{Zf}$, $\dgz$, $\dkapa$, $\lambda_\gamma$, $c_{3G}$ can be read off from the equations above, and are listed in table~\ref{tab:hb}. Note that we have defined $[\dgL^{Wq}]_{ij}$ in the gauge eigenstate basis, as opposed to the current version of~\cite{HiggsBasis} where it is defined in the mass eigenstate basis. The coupling relation
\beq
\dkapz = \dgz-\frac{\sw^2}{\cw^2}\dkapa
\eeqn
holds as in general nonuniversal theories. It is clear from table~\ref{tab:hb} that among the 4 oblique parameters $\hat S, \hat T, W, Y$ in the electroweak sector, only 3 linear combinations $\De_{1,2,3}$ enter the Higgs basis couplings discussed above. It is well-known that the fourth independent oblique parameter is accessible only through off-$Z$-pole four-fermion processes, such as $e^+e^-\to\bar f f$ at LEP2~\cite{Barbieri:2004qk}. In the Higgs basis, the contributing parameters are coefficients of 4-fermion operators, which we collectively denote by $c_{4f}$. They are linear combinations of $W, Y$ [see \eqref{add4f}], and, if we go beyond the electroweak sector, also $Z, c_{2y}$. On the other hand, the $W^\pm$ coupling with right-handed quarks $\dgR^{Wq}$, and  the dipole-type couplings $d_{Vf}$ are not present in universal theories at tree level.

Finally, we look at the Higgs sector. The Higgs boson kinetic term in \eqref{Lutc} already satisfies the first Higgs basis defining condition, so $\bar h=\hat h$. To preserve the leading-order expressions of the Higgs boson and SM fermion masses
\beq
m_H^{\text{LO}} = \sqrt{2\bar\lambda}\bar v = \sqrt{2\hat\lambda}\hat v,\quad m_{f'}^{\text{LO}} = \frac{\bar y_{f'}\bar v}{\sqrt{2}} = \frac{\hat y_{f'}\hat v}{\sqrt{2}},
\eeqn
as required by the second Higgs basis defining condition, we should, by \eqref{vredef-hb}, have
\beq
\bar\lambda = (1+\De_2)\hat\lambda,\quad \bar y_{f'} = \Bigl(1+\frac{\De_2}{2}\Bigr)\hat y_{f'}.
\eeqn
It follows that the triple-Higgs and Higgs-fermion interactions become
\bseq
\beqa
\L_{h^3} &=& -\Bigl(1+\Delta\kappa_3-\frac{\De_2}{2}\Bigr) \bar\lambda\bar v\bar h^3 = -(1+\Delta\kappa_3)\hat\lambda\hat v\hat h^3 \equiv -(\hat\lambda+\delta\lambda_3)\hat v\hat h^3, \\
\L_{hff} &=& -\Bigl(1+\DkapFb-\frac{\De_2}{2}\Bigr)\sum_{f'}\frac{\bar y_{f'}}{\sqrt{2}}\bar h \bar f' f' = -(1+\DkapFb)\sum_{f'}\frac{\hat y_{f'}}{\sqrt{2}}\hat h \bar f' f' \CR
&\equiv& -\sum_{f'} \bigl(\delta_{ij}+[\delta y_{f'}]_{ij}\bigr)\frac{\hat y_{f'}}{\sqrt{2}}\hat h \bar f'_i f'_j,
\eeqan
\eseqn
from which one can read off the Higgs basis couplings $\delta\lambda_3$ and $[\delta y_{f'}]_{ij}$; see table~\ref{tab:hb}.

To derive the Higgs-vector boson couplings, further field redefinitions, or equivalently, applications of EoM, are needed. We see from \eqref{LCC} and \eqref{LNC} that anomalous $Vff$ couplings have been generated, but not accompanied by $hVff, h^2Vff$ vertices. To generate the latter with coefficients required by the third Higgs basis defining condition, we reorganize the anomalous $Vff$ interaction terms and apply the EoM as follows,
\bseq
\beqa
&&\sum_{f=q,l}[\dgL^{Wf}]_{ij}\frac{g}{\sqrt{2}}(W^+_\mu\bar f_i\gamma^\mu\sigma^+f_j +\text{h.c.}) \CR
&&\quad = \sum_{f=q,l}[\dgL^{Wf}]_{ij}\Bigl(1+\frac{h}{v}\Bigr)^2\frac{g}{\sqrt{2}}(W^+_\mu\bar f_i\gamma^\mu\sigma^+f_j +\text{h.c.}) +\Delta\L_{hW}, \\
&&\sum_f [\delta g_{L/R}^{Zf}]_{ij}\frac{e}{\cw\sw}Z_\mu\bar f_i\gamma^\mu f_j = \sum_f [\delta g_{L/R}^{Zf}]_{ij} \Bigl(1+\frac{h}{v}\Bigr)^2\frac{e}{\cw\sw}Z_\mu\bar f_i\gamma^\mu f_j +\Delta\L_{hZ},
\eeqan
\eseqn
where
\bseq
\beqa
\Delta\L_{hW} &=& -2\Bigl(\frac{h}{v}+\frac{h}{2v^2}\Bigr)\sum_{f=q,l}[\dgL^{Wf}]_{ij}\frac{g}{\sqrt{2}}(W^+_\mu\bar f_i\gamma^\mu\sigma^+f_j +\text{h.c.}) \CR
&=& -\sqrt{2}\Bigl(\frac{h}{v}+\frac{h}{2v^2}\Bigr)\Bigl(\frac{\cw^2}{\cw^2-\sw^2}\frac{\De_1}{2} -\frac{\sw^2}{\cw^2-\sw^2}\De_3\Bigr) \bigl[W^+_\mu(J_W^{1\mu}+iJ_W^{2\mu})+\text{h.c.}\bigr] \CR
&\xrightarrow{\EoM}& -\sqrt{2}\Bigl(\frac{h}{v}+\frac{h}{2v^2}\Bigr)\Bigl(\frac{\cw^2}{\cw^2-\sw^2}\frac{\De_1}{2} -\frac{\sw^2}{\cw^2-\sw^2}\De_3\Bigr) \CR
&&\quad \bigl[W^+_\mu(D_\nu W^{1\mu\nu}+iD_\nu W^{2\mu\nu} -igH^\dagger\sigma^+\Dlr_\mu H) +\text{h.c.}\bigr] \CR
&=& \Bigl(\frac{h}{v}+\frac{h}{2v^2}\Bigr) \Bigl(\frac{\cw^2}{\cw^2-\sw^2}\frac{\De_1}{2} -\frac{\sw^2}{\cw^2-\sw^2}\De_3\Bigr) \CR
&&\quad \biggl[4\Bigl(\frac{gv}{2}\Bigr)^2W^+_\mu W^{-\mu}\Bigl(1+\frac{h}{v}\Bigr)^2 -2(W^-_\mu\partial_\nu W^{+\mu\nu}+\text{h.c.}) +\O(V^3) \biggr], \\
\Delta\L_{hZ} &=& -2\Bigl(\frac{h}{v}+\frac{h}{2v^2}\Bigr)\sum_f [\delta g_{L/R}^{Zf}]_{ij}\frac{e}{\cw\sw}Z_\mu\bar f_i\gamma^\mu f_j \CR
&=& -\frac{2e}{\cw\sw(\cw^2-\sw^2)}\Bigl(\frac{h}{v}+\frac{h}{2v^2}\Bigr)\sum_f \Bigl[(\cw^2T^3_f+\sw^2Y_f)\frac{\De_1}{2} -\sw^2Q_f\De_3\Bigr] Z_\mu\bar f_i\gamma^\mu f_i \CR
&=& -\frac{2}{\cw^2-\sw^2}\Bigl(\frac{h}{v}+\frac{h}{2v^2}\Bigr)Z_\mu \Bigl[(\cw J_W^{3\mu}+\sw J_B^\mu)\frac{\De_1}{2}-\frac{\sw}{\cw}J_{\text{EM}}^\mu\De_3\Bigr] \CR
&\xrightarrow{\EoM}& -\frac{2}{\cw^2-\sw^2}\Bigl(\frac{h}{v}+\frac{h}{2v^2}\Bigr)Z_\mu \biggl\{\Bigl[\cw D_\nu W^{3\mu\nu}+\sw \partial_\nu B^{\mu\nu} \CR
&&\qquad -\frac{ie}{2\cw\sw}H^\dagger(\cw^2\sigma^3+\sw^2)\Dlr^\mu H\Bigr]\frac{\De_1}{2} -\frac{\sw}{\cw}\partial_\nu A^{\mu\nu}\De_3\biggr\} \CR
&=& \Bigl(\frac{h}{v}+\frac{h}{2v^2}\Bigr) \biggl[\De_1\Bigl(\frac{gv}{2\cw}\Bigr)^2 Z_\mu Z^\mu\Bigl(1+\frac{h}{v}\Bigr)^2 -\De_1 Z^\mu\partial^\nu Z_{\mu\nu} \CR
&&\qquad -\frac{2\cw\sw}{\cw^2-\sw^2}\Bigl(\De_1-\frac{\De_3}{\cw^2}\Bigr) Z^\mu\partial^\nu A_{\mu\nu} +\O(V^3) \biggr].
\eeqan
\eseqn
One can then add $\Delta\L_{hW},\Delta\L_{hZ}$ to the Higgs-vector boson interactions in \eqref{Lutc} [with the replacements \eqref{replace}], and apply the redefinitions \eqref{ZAredef}, \eqref{Oinredef}, \eqref{Oinredef2}. For example, the zero-derivative $hZZ$ coupling reads
\beqa
&& \Bigl(1+\DkapVb-\frac{3\De_2}{2}\Bigr)(1-2\De_1+2\De_2)\frac{\bar h}{\bar v}\Bigl(\frac{\bar g\bar v}{2\cwb}\Bigr)^2\bar Z_\mu\bar Z^\mu +\De_1\frac{h}{v}\Bigl(\frac{gv}{2\cw}\Bigr)^2 Z_\mu Z^\mu \CR
&=& \Bigl(1+\DkapVb-\frac{3\De_2}{2}\Bigr)(1-2\De_1+2\De_2) \Bigl(1+\De_1-\frac{\De_2}{2}\Bigr)\frac{\hat h}{\hat v}\Bigl(\frac{\hat g\hat v}{2\cwh}\Bigr)^2\hat Z_\mu\hat Z^\mu \CR
&& +\De_1\frac{\hat h}{\hat v}\Bigl(\frac{\hat g\hat v}{2\cwh}\Bigr)^2 \hat Z_\mu \hat Z^\mu \CR
&=& (1+\DkapVb)\frac{\hat h}{\hat v}\Bigl(\frac{\hat g\hat v}{2\cwh}\Bigr)^2\hat Z_\mu\hat Z^\mu \equiv(1+\delta c_z)\frac{\hat h}{\hat v}\Bigl(\frac{\hat g\hat v}{2\cwh}\Bigr)^2\hat Z_\mu\hat Z^\mu, 
\eeqan
so that the Higgs basis coupling $\delta c_z=\DkapVb$. Similarly, one can work out the zero-derivative $hWW$ coupling, and show explicitly the coupling relation
\beq
\delta c_w = \delta c_z+4\delta_m,
\eeqn
which holds at the dimension-6 level in general nonuniversal theories. On the other hand, the above procedure does not affect the terms in \eqref{Lutc} proportional to $f_{vv'}$, so the latter are directly identified with the Higgs basis parameters $c_{vv'}$. Other parameters in the Higgs sector, including $c_{v\square}$, and couplings of 2 Higgs bosons to fermions or vector bosons, can also be derived by this procedure. We have explicitly checked that they satisfy the generally-valid coupling relations listed in~\cite{HiggsBasis}.

\begin{table}[tbp]
\centering
\begin{tabular}{|l|l|}
\hline
Higgs basis coupling & universal parameters \\
\hline
$\delta_m$ & $\frac{\cw^2}{\cw^2-\sw^2}\frac{\De_1}{2} -\frac{\De_2}{2} -\frac{\sw^2}{\cw^2-\sw^2}\De_3$ \\
\hline
$[\dgL^{Wf}]_{ij}\,\, (f=q,l)$ & $\delta_{ij}\bigl(\frac{\cw^2}{\cw^2-\sw^2}\frac{\De_1}{2} -\frac{\sw^2}{\cw^2-\sw^2}\De_3\bigr)$ \\
$[\dgL^{Zf}]_{ij}\,\, (f=u_L,d_L,e_L,\nu)$ & $\delta_{ij}\Bigl[T^3_f\frac{\De_1}{2} +Q_f\frac{\sw^2}{\cw^2-\sw^2}\bigl(\frac{\De_1}{2}-\De_3\bigr)\Bigr]$ \\
$[\dgR^{Zf}]_{ij}\,\, (f=u_R,d_R,e_R)$ & $\delta_{ij}Q_f\frac{\sw^2}{\cw^2-\sw^2}\bigl(\frac{\De_1}{2}-\De_3\bigr)$ \\
\hline
$\dgz$ & $\Dgzb -\frac{\De_2}{\cw^2} +\frac{\sw^2}{\cw^2-\sw^2}\bigl(\frac{\De_1}{2\sw^2}-\frac{\De_3}{\cw^2}\bigr)$ \\
$\dkapa$ & $\Dkapab$ \\
$\lambda_\gamma$ & $\lamab$ \\
$c_{3G}$ & $-\frac{2}{3g_s^2g^2}\lamgb$ \\
\hline
$\delta\lambda_3$ & $\lambda\Delta\kappa_3$ \\
$[\delta y_{f'}]_{ij}\,\, (f'=u,d,e)$ & $\delta_{ij}\DkapFb$ \\
$\delta c_z$ & $\DkapVb$ \\
\hline
$c_{gg}, c_{z\gamma}, c_{\gamma\gamma}$ & $f_{gg}, f_{z\gamma}, f_{\gamma\gamma}$, respectively \\
\hline
$c_{4f}$ & combinations of $W, Y, Z, c_{2y}$ \\
\hline
$[\dgR^{Wq}]_{ij}, [d_{Vf}]_{ij}$ & 0 \\
\hline
\end{tabular}
\caption{\label{tab:hb} Higgs basis couplings in terms of the universal parameters. $\De_{1,2,3}$ are independent linear combinations of $\hat S, \hat T, W, Y$ defined in \eqref{Dedef}. $c_{4f}$ collectively denotes four-fermion effective couplings, and $d_{Vf}$ stands for the dipole-type $Vff$ couplings.}
\end{table}

Table~\ref{tab:hb} summarizes the Higgs basis couplings expressed in terms of the universal parameters found in this subsection. The Higgs basis couplings listed in the first column of the table actually constitute a complete basis of independent couplings modulo two redundancies
\beq
\dgL^{Z\nu} = \dgL^{Ze}+\dgL^{Wl},\quad \dgL^{Wq} = \dgL^{Zu}-\dgL^{Zd},
\eeq{couprel-ew}
which are among the generally-valid coupling relations in~\cite{HiggsBasis}. The set of independent couplings chosen here differs slightly from that in~\cite{HiggsBasis}, in that two of the $hVV$ couplings $c_{zz}, c_{z\square}$ have been traded for the anomalous TGCs $\dgz, \dkapa$. Some of the coupling relations listed in~\cite{HiggsBasis} take a slightly different (and simpler) form when $\dgz, \dkapa$ are used as independent couplings in place of $c_{zz}, c_{z\square}$:
\bseq
\beqa
c_{ww} &=& c_{z\gamma}+\sw^2c_{\gamma\gamma}+\frac{2}{g^2}\dkapa, \\
c_{zz} &=& (\cw^2-\sw^2)c_{z\gamma} +\cw^2\sw^2c_{\gamma\gamma} +\frac{2}{g^2}\dkapa, \\
c_{w\square} &=& -\frac{2\cw^2}{g^2}\dgz, \\
c_{z\square} &=& -\frac{2}{g^2}\Bigl[(\cw^2-\sw^2)\dgz +\frac{\sw^2}{\cw^2}\dkapa\Bigr], \\
c_{\gamma\square} &=& -\frac{2}{g^2}(2\cw^2\dgz-\dkapa).
\eeqan
\eseq{depcoup}
From these equations it is clear that new physics contributions to the Higgs-vector boson couplings are related to the anomalous TGCs, a fact that has been used recently to extract the TGC parameters from Higgs data~\cite{Corbett:2013pja,Falkowski:2015jaa}. This connection will be demonstrated in more detail with an example in section~\ref{sec:obs-tgc}.

\subsection{Universal effects in universal theories}
\label{sec:hb-cor}

Table~\ref{tab:hb} shows the following special features of universal theories at leading order.
\begin{itemize}
\item All the $Vff$ vertex corrections are determined by only 2 parameters $\De_1, \De_3$. Focusing on one generation for simplicity, we can write down 5 relations among the 7 independent couplings $\dgL^{Ze}, \dgR^{Ze}, \dgL^{Wl}, \dgL^{Zu}, \dgR^{Zu}, \dgL^{Zd}, \dgR^{Zd}$:
\beqa
&& \dgL^{Wq} = \dgL^{Wl},\quad \frac{\dgR^{Zu}}{Q_u} = \frac{\dgR^{Zd}}{Q_d} =  \frac{\dgR^{Ze}}{Q_e}, \CR
&& \dgL^{Ze}+\dgL^{Z\nu} = \dgR^{Ze},\quad \dgL^{Zu}+\dgL^{Zd} = \dgR^{Zu}+\dgR^{Zd}.
\eeqa{urew}
\item All the $hff$ vertices are rescaled by a common factor $(1+\DkapFb)$ compared to the SM ones, i.e.\
\beq
[\delta y_u]_{ij} = [\delta y_d]_{ij} = [\delta y_e]_{ij} = \delta_{ij}\DkapFb.
\eeq{ury}
\item The plethora of four-fermion couplings are all linear combinations of 4 parameters $W, Y, Z, c_{2y}$.
\item The independent couplings $\dgR^{Wq}$ and $d_{Vf}$ are not generated.
\end{itemize}
These features actually provide another way to define universal theories, by clarifying the sense in which the indirect new physics effects are ``universal.'' All of them are restrictions on the way in which the SM fermions couple, which originate from the statement of universal theories definition in section~\ref{sec:def-gen}. In particular, the relations shown in \eqref{urew} and \eqref{ury} restrict the patterns of electroweak and Yukawa coupling modifications in universal theories at leading order (these patterns will be slightly distorted by RG evolution~\cite{followup}; see also~\cite{Elias-Miro:2013mua}). The bosonic sector, on the other hand, has the same number of independent couplings in universal and nonuniversal theories: $\delta_m$, $\dgz$, $\dkapa$, $\lambda_\gamma$, $c_{3G}$, $\delta\lambda_3$, $\delta c_z, c_{gg}$, $c_{z\gamma}$, $c_{\gamma\gamma}$. These 10 independent couplings among SM bosons, plus the 6 additional independent couplings involving SM fermions (2 for $Vff$, 1 for $hff$, and 3 more for $4f$), give the correct number of independent parameters (16) in universal theories.

To close this section, we remark that while universal and nonuniversal theories have often been discussed in different languages (e.g.\ oblique vs.\ vertex corrections), and argued to be more conveniently analyzed in different SMEFT bases (see e.g.~\cite{Elias-Miro:2013gya,Elias-Miro:2013mua,Biekoetter:2014jwa}), the former is really a limit of the latter. This seemingly trivial but perhaps less appreciated (from the EFT perspective) point is made clear in this section, as we have seen how the limit can be explicitly taken in the Higgs basis framework. The special features of universal theories in this limit listed above distinguish them from the more general nonuniversal theories.

\section{From universal parameters to observables}
\label{sec:obs}

\subsection{Precision electroweak observables}
\label{sec:obs-ew}
In~\cite{Wells:2014pga}, we demonstrated that, with the knowledge of the Higgs boson mass, precision electroweak analyses can be formulated in terms of expansion formulas, taking into account both the state-of-the-art SM calculations and perturbative new physics corrections. One interesting example shown in~\cite{Wells:2014pga} is BSM scenarios where the new particles affect precision electroweak observables predominantly via contributions to the vector boson self-energies. For the $Z$-pole observables and $m_W$, only 6 quantities enter the calculations,
\beqa
&& \pi_{ww}^0 \equiv \frac{\Pi_{WW}(0)}{m_W^2},\quad \pi_{ww} \equiv \frac{\Pi_{WW}(m_W^2)}{m_W^2},\quad \pi_{zz} \equiv \frac{\Pi_{ZZ}(m_Z^2)}{m_Z^2}, \CR
&& \pi'_{zz} \equiv \Pi'_{ZZ}(m_Z^2),\quad \pi_{\gamma z} \equiv \frac{\Pi_{\gamma Z}(m_Z^2)}{m_Z^2},\quad \pi'_{\gamma\gamma} \equiv \Pi'_{\gamma\gamma}(0).
\eeqa{pidef}
The fractional shifts of the observables due to new physics, defined as
\beq
\dbNP\Obs_i \equiv \frac{\Obs_i-\Obs_i^\SM}{\Obs_i^\SM},
\eeq{dbOdef}
are given at LO by
\beq
\dbNP\Obs_i = b^0_{i,ww}\pi^0_{ww} +b_{i,ww}\pi_{ww} +b_{i,zz}\pi_{zz} +b'_{i,zz}\pi'_{zz} +b_{i,\gamma z}\pi_{\gamma z} +b'_{i,\gamma\gamma}\pi'_{\gamma\gamma},
\eeq{bdef}
with the $b$-coefficients tabulated in~\cite{Wells:2014pga}. The $\pi$-parameters here include only the new physics contributions, and correspond to $\pi^\NP$ in~\cite{Wells:2014pga}; $\Pi_{VV'}(p^2)$ is defined in \eqref{Pidef}. 

These results do not rely on the SMEFT framework, and are valid in complete generality. But since the BSM scenarios under consideration are by assumption universal theories, it is useful to recast \eqref{bdef} in terms of the universal parameters $\hat S$, $\hat T$, $W$, $Y$ (the fifth oblique parameter $Z$ is not relevant here since we focus on observables in the electroweak sector) when the effective Lagrangian is truncated at dimension 6. Using the results in section~\ref{sec:utc-obl}, we find, after the field and parameter redefinitions necessary to satisfy the oblique parameters defining conditions [i.e.\ replacing $\Pi_{VV}$ by $\bar\Pi_{VV}$ in \eqref{pidef}],
\beqa
&& \pi^0_{ww} = 0,\quad \pi_{ww} = -W,\quad \pi_{zz} = 2\sw^2\hat S -\hat T -W -\frac{\sw^2}{\cw^2}Y, \CR
&& \pi'_{zz} = 2\Bigl(\sw^2\hat S -W -\frac{\sw^2}{\cw^2}Y\Bigr),\,\, \pi_{\gamma z} = -\frac{\sw}{\cw}\bigl[(\cw^2-\sw^2)\hat S +W -Y\bigr],\,\, \pi'_{\gamma\gamma} = -2\sw^2\hat S.
\eeqa{pi-utc}
These equations were previously worked out in~\cite{Ellis:2014jta} in the special case $W=Y=0$.

To take one step further, we note that \eqref{bdef} is actually a redundant representation of $\dbNP\Obs_i$. There are 3 relations among the 6 $b$-coefficients, associated with the 3 flat directions in the space of the 6 $\pi$-parameters, along which observables do not change. They can be found by rescaling the SM parameters and fields such that all the new physics effects on the electroweak observables are still captured by the 6 $\pi$-parameters. Such rescalings cannot change the observables (when they are expressed in terms of input observables), but shift the $\pi$-parameters along the flat directions:
\begin{itemize}
\item $g\to(1+\frac{\delta}{2})g$, $W_\mu^a\to(1-\frac{\delta}{2})W_\mu^a$ $\Rightarrow$ $\Delta\boldsymbol{\pi}=(0,1,\cw^2,\cw^2,\cw\sw,\sw^2)\delta$;
\item $g'\to(1+\frac{\delta}{2})g'$, $B_\mu\to(1-\frac{\delta}{2})B_\mu$ $\Rightarrow$ $\Delta\boldsymbol{\pi}=(0,0,\sw^2,\sw^2,-\cw\sw,\cw^2)\delta$;
\item $v\to(1+\frac{\delta}{2})v$ $\Rightarrow$ $\Delta\boldsymbol{\pi}=(1,1,1,0,0,0)\delta$.
\end{itemize}
Here $\boldsymbol{\pi}\equiv(\pi^0_{ww},\pi_{ww},\pi_{zz},\pi'_{zz},\pi_{\gamma z},\pi'_{\gamma\gamma})$, and $\Delta\boldsymbol{\pi}$ denotes the shift in $\boldsymbol{\pi}$. We can directly read off the relations among the $b$-coefficients that must be satisfied,
\bseq
\beqa
b_{ww} +\cw^2(b_{zz}+b'_{zz}) +\cw\sw b_{\gamma z} +\sw^2 b'_{\gamma\gamma} &=& 0, \\
\sw^2(b_{zz}+b'_{zz}) -\cw\sw b_{\gamma z} +\cw^2 b'_{\gamma\gamma} &=& 0, \\
b^0_{ww} +b_{ww} +b_{zz} &=& 0.
\eeqan
\eseqn
It is clear from the calculations and numerical results in~\cite{Wells:2014pga} that these relations indeed hold. They allow us to eliminate 3 of the 6 $b$-coefficients, which we choose to be $b^0_{ww}$, $b_{zz}$, $b'_{\gamma\gamma}$ for illustration. Eq.~\eqref{bdef} then becomes
\beqa
\dbNP\Obs_i &=& b_{ww}\Bigl[\pi_{ww} -\frac{\cw^2}{\cw^2-\sw^2}\pi_{zz} +\frac{\sw^2}{\cw^2-\sw^2}(\pi^0_{ww}+\pi'_{\gamma\gamma})\Bigr] +b'_{zz}(\pi'_{zz}-\pi_{zz}+\pi^0_{ww}) \CR
&& +b_{\gamma z}\Bigl[\pi_{\gamma z} +\frac{\cw\sw}{\cw^2-\sw^2}(\pi^0_{ww}+\pi'_{\gamma\gamma}-\pi_{zz})\Bigr] \CR
&=& \Bigl(\frac{\cw^2}{\cw^2-\sw^2}b_{ww} +\frac{\cw\sw}{\cw^2-\sw^2}b_{\gamma z} +b'_{zz}\Bigr)\De_1 -b_{ww}\De_2 \CR
&& -\frac{1}{\cw^2-\sw^2}\Bigl(\frac{\sw}{\cw}b_{\gamma z} +2\sw^2b_{ww}\Bigr)\De_3,
\eeqan
where we have used \eqref{pi-utc} to arrive at the second equation. As expected, the result depends on the 4 oblique parameters $\hat S$, $\hat T$, $W$, $Y$ only through the 3 linear combinations $\De_{1,2,3}$, defined in \eqref{Dedef}. This is a well-known fact~\cite{Altarelli:1990zd,Altarelli:1991fk,Barbieri:2004qk}, and is also obvious from the values of the Higgs basis parameters in table~\ref{tab:hb}.

\subsection{Interplay between $e^+e^-\to W^+W^-$ and $h\to Z\ell^+\ell^-$}
\label{sec:obs-tgc}

There has been quite some interest recently in the interplay between TGC measurements and Higgs data~\cite{Corbett:2013pja,Falkowski:2015jaa} (see also~\cite{Pomarol:2013zra,Ellis:2014jta}). As we have seen in section~\ref{sec:hb-c}, the relevant Higgs basis couplings are correlated. The measurements of the TGCs are currently dominated by $e^+e^-\to W^+W^-$ at LEP2, for which an EFT calculation in the case of universal theories has been presented in~\cite{Wells:2015eba}. On the other hand, measurement of the spectrum of $h\to Z\ell^+\ell^-$, a very clean decay channel, will be sensitive to an overlapping set of SMEFT parameters. The calculation of this process has been recently discussed in~\cite{Isidori:2013cla,Grinstein:2013vsa,Buchalla:2013mpa,Beneke:2014sba} (see also~\cite{Gonzalez-Alonso:2014eva,Gonzalez-Alonso:2015bha,Bordone:2015nqa}). Here we recast this calculation in the Higgs basis framework, and map the results to universal parameters in the case of universal theories. This will provide an illustration of the Higgs basis at work, and help address the concerns raised in~\cite{Trott:2014dma} regarding theory consistency related to the defining assumptions of the $S$ parameter and anomalous TGCs.

To begin with, we specify the notation and kinematics. We label the final state particles $Z$, $\ell^+$, $\ell^-$ by $1,2,3$, respectively, with $p_1^\mu$, $p_2^\mu$, $p_3^\mu$ being the corresponding 4-momenta. We denote the invariant mass squared of two particles by $m_{ij}^2=(p_i+p_j)^2$, and define $q^\mu=p_2^\mu+p_3^\mu$ so that $q^2=m_{23}^2$. The initial-state $h$ and the final-state $Z$ will be assumed on-shell, and lepton masses will be neglected. We will be interested in the differential decay rate $\frac{d\Gamma}{dq^2}$ for either $\ell=e$ or $\ell=\mu$ or $\ell=\tau$, with the polarizations of $Z$ and the chiralities of $\ell^+\ell^-$ summed over [$\ell$ should not be confused with the $SU(2)_L$ doublet field $l$]. We have,
\beq
\frac{d\Gamma}{dq^2} = \frac{1}{256\pi^3m_h^3}\int_{m_-^2}^{m_+^2}|\M|^2dm_{12}^2,
\eeq{intm12}
where
\beq
m_\pm^2 = \frac{1}{2}\biggl[m_h^2+m_Z^2-q^2 \pm\sqrt{q^4-2q^2(m_h^2+m_Z^2)+(m_h^2-m_Z^2)^2}\biggr].
\eeqn

To calculate $|\M|^2$, the matrix element squared with the final state polarizations and chiralities summed over as specified above, we need the following interaction terms in the Higgs basis Lagrangian,
\beqa
\L &\supset& \frac{h}{v}\Bigl[(1+\delta c_z)\Bigl(\frac{gv}{2\cw}\Bigr)^2Z_\mu Z^\mu +c_{zz}\frac{g^2}{4\cw^2}Z_{\mu\nu}Z^{\mu\nu} +c_{z\gamma}\frac{gg'}{2}Z_{\mu\nu}A^{\mu\nu} \CR
&&\quad +c_{z\square}g^2Z_\mu\partial_\nu Z^{\mu\nu} +c_{\gamma\square}gg'Z_\mu\partial_\nu A^{\mu\nu} \Bigr] \CR
&& +\frac{g}{\cw}Z_\mu\sum_{f=\ell}\bigl[(g_L+\dgL^{Zf})\bar f_L\gamma^\mu f_L +(g_R+\dgR^{Zf})\bar f_R\gamma^\mu f_R\Bigr] \CR
&& +\frac{2g}{\cw}\frac{h}{v}Z_\mu\sum_{f=\ell}(\dgL^{hZf}\bar f_L\gamma^\mu f_L +\dgR^{hZf}\bar f_R\gamma^\mu f_R).
\eeqan
We have dropped the hats on the fields and parameters for simplicity, and defined 
\beq
\{g_L,\, g_R\} = \bigl\{T^3_f-Q_f\sw^2,\, -Q_f\sw^2\bigr\} = \Bigl\{-\frac{1}{2}+\sw^2,\, \sw^2\Bigr\}\quad\text{for}\,\, f=\ell.
\eeqn
In the SM, $h\to Z\ell^+\ell^-$ proceeds through the single diagram $h\to ZZ^*\to Z\ell^+\ell^-$ at LO. Besides corrections to the vertices in this diagram, there are two additional LO diagrams, $h\to Z\gamma^*\to Z\ell^+\ell^-$ and $h\to Z\ell^+\ell^-$ (via the 4-point vertex), in the SMEFT. We find, up to loop corrections and higher order terms in $\frac{v^2}{\Lambda^2}$,
\beqa
|\M|^2 &=& \frac{g^4}{\cw^4} \biggl\{(1+\delta c_z)^2\bigl[(g_L+\dgL^{Zf})^2+(g_R+\dgR^{Zf})^2\bigr]\CR
&&\quad \frac{m_Z^2(2q^2-m_h^2)+m_{12}^2(m_h^2+m_Z^2-q^2)-m_{12}^4}{(q^2-m_Z^2)^2} \CR
&& +\Bigl[c_{zz}\frac{g^2}{\cw^2}(g_L^2+g_R^2)\frac{q^2}{q^2-m_Z^2} +c_{z\gamma}e^2Q(g_L+g_R)\Bigr]\frac{q^2+m_Z^2-m_h^2}{q^2-m_Z^2} \CR
&& +\Bigl[c_{z\square}g^2(g_L^2+g_R^2)\frac{q^2+m_Z^2}{q^2-m_Z^2}+c_{\gamma\square}e^2Q(g_L+g_R) +2(g_L\dgL^{hZf}+g_R\dgR^{hZf})\Bigr]\CR
&&\quad \frac{m_Z^2(2q^2-m_h^2)+m_{12}^2(m_h^2+m_Z^2-q^2)-m_{12}^4}{m_Z^2(q^2-m_Z^2)} \biggr\},
\eeqan
where $Q=-1$. The contribution from each diagram is apparent from this expression. Integrating over $m_{12}^2$ as in \eqref{intm12}, we get,
\beqa
\frac{d\Gamma}{dq^2} &=& \frac{g_L^2+g_R^2}{1536\pi^3}\frac{g^4}{\cw^4} \frac{\sqrt{q^4-2q^2(m_h^2+m_Z^2)+(m_h^2-m_Z^2)^2}}{m_h^3} \CR
&&\quad \frac{q^4-2q^2(m_h^2-5m_Z^2)+(m_h^2-m_Z^2)^2}{(q^2-m_Z^2)^2} \Bigl(1 +\dbNP\frac{d\Gamma}{dq^2}\Bigr),
\eeqan
where the fractional shift due to new physics, defined in \eqref{dbOdef}, is given by
\beqa
\dbNP\frac{d\Gamma}{dq^2} &=& 2\delta c_z +\frac{2g_L}{g_L^2+g_R^2}\Bigl(\dgL^{Zf}+\frac{q^2-m_Z^2}{m_Z^2}\dgL^{hZf}\Bigr) +\frac{2g_R}{g_L^2+g_R^2}\Bigl(\dgR^{Zf}+\frac{q^2-m_Z^2}{m_Z^2}\dgR^{hZf}\Bigr) \CR
&& +\frac{6q^2(q^2+m_Z^2-m_h^2)}{q^4-2q^2(m_h^2-5m_Z^2)+(m_h^2-m_Z^2)^2} \Bigl[\frac{g^2}{\cw^2}c_{zz} +\frac{Q(g_L+g_R)}{g_L^2+g_R^2}\frac{q^2-m_Z^2}{q^2}e^2c_{z\gamma}\Bigr]\CR
&& +\frac{q^2+m_Z^2}{m_Z^2}g^2c_{z\square} +\frac{Q(g_L+g_R)}{g_L^2+g_R^2}\frac{q^2-m_Z^2}{m_Z^2}e^2c_{\gamma\square}.
\eeqan
Using $\delta g_{L,R}^{hZf}=\delta g_{L,R}^{Zf}$ and \eqref{depcoup} to eliminate $\delta g_{L,R}^{hZf}$, $c_{zz}$, $c_{z\square}$, $c_{\gamma\square}$, we can write the result in terms of the independent couplings,
\beqa
\dbNP\frac{d\Gamma}{dq^2} &=& 2\delta c_z +\frac{2}{g_L^2+g_R^2}\frac{q^2}{m_Z^2}(g_L\dgL^{Zf} +g_R\dgR^{Zf}) \CR
&& +\frac{6q^2(q^2+m_Z^2-m_h^2)}{q^4-2q^2(m_h^2-5m_Z^2)+(m_h^2-m_Z^2)^2}e^2 \CR
&&\qquad \biggl\{\Bigl[\frac{\cw^2-\sw^2}{\cw^2\sw^2} +\frac{Q(g_L+g_R)}{g_L^2+g_R^2}\frac{q^2-m_Z^2}{q^2}\Bigr] c_{z\gamma} +c_{\gamma\gamma}\biggr\} \CR
&& -2\Bigl[(\cw^2-\sw^2)\frac{q^2+m_Z^2}{m_Z^2} +2\cw^2\sw^2\frac{Q(g_L+g_R)}{g_L^2+g_R^2}\frac{q^2-m_Z^2}{m_Z^2}\Bigr]\dgz \CR
&& +\frac{2}{\cw^2}\Bigl[\frac{6q^2(q^2+m_Z^2-m_h^2)}{q^4-2q^2(m_h^2-5m_Z^2)+(m_h^2-m_Z^2)^2} \CR
&&\qquad -\sw^2\frac{q^2+m_Z^2}{m_Z^2} +\cw^2\sw^2\frac{Q(g_L+g_R)}{g_L^2+g_R^2}\frac{q^2-m_Z^2}{m_Z^2}\Bigr]\dkapa.
\eeqa{dGam}

Up to now, our calculation has been completely general, and is valid also for nonuniversal theories. Specializing to the case of universal theories, we can use table~\ref{tab:hb} to rewrite \eqref{dGam} in terms of the universal parameters $\De_{1,2,3}$ (combinations of $\hat S$, $\hat T$, $W$, $Y$), $\Dgzb$, $\Dkapab$, $\DkapVb$, $f_{z\gamma}$, $f_{\gamma\gamma}$. In this case, precision electroweak measurements constrain the oblique parameters $\De_{1,2,3}$ to be very small. In the limit where these parameters vanish,
\beqa
\dbNP\frac{d\Gamma}{dq^2} &=& -2\Bigl[(\cw^2-\sw^2)\frac{q^2+m_Z^2}{m_Z^2} +2\cw^2\sw^2\frac{Q(g_L+g_R)}{g_L^2+g_R^2}\frac{q^2-m_Z^2}{m_Z^2}\Bigr]\Dgzb \CR
&& +\frac{2}{\cw^2}\Bigl[\frac{6q^2(q^2+m_Z^2-m_h^2)}{q^4-2q^2(m_h^2-5m_Z^2)+(m_h^2-m_Z^2)^2} \CR
&&\qquad -\sw^2\frac{q^2+m_Z^2}{m_Z^2} +\cw^2\sw^2\frac{Q(g_L+g_R)}{g_L^2+g_R^2}\frac{q^2-m_Z^2}{m_Z^2}\Bigr]\Dkapab \CR
&& +2\DkapVb +\frac{6q^2(q^2+m_Z^2-m_h^2)}{q^4-2q^2(m_h^2-5m_Z^2)+(m_h^2-m_Z^2)^2}e^2 \CR
&&\qquad \biggl\{\Bigl[\frac{\cw^2-\sw^2}{\cw^2\sw^2} +\frac{Q(g_L+g_R)}{g_L^2+g_R^2}\frac{q^2-m_Z^2}{q^2}\Bigr] f_{z\gamma} +f_{\gamma\gamma}\biggr\} \quad (\De_{1,2,3}\to0).
\eeqa{dGam-u}
The dependence on the anomalous TGC parameters $\Dgzb$, $\Dkapab$ can be clearly seen from this equation. 
The same parameters enter the $e^+e^-\to W^+W^-$ observables in the same limit $\De_{1,2,3}\to0$. For example, translating the results in~\cite{Wells:2015eba} into the parameterizations in this paper, we find that, at $\sqrt{s}=200$~GeV, the unpolarized cross section is shifted by
\beq
\dbNP\sigma = -0.0374\Dgzb -0.0960\Dkapab -0.0537\lamab \quad (\De_{1,2,3}\to0).
\eeq{sig-eeww}
Therefore, the anomalous TGC parameters $\Dgzb$, $\Dkapab$ extracted from $e^+e^-\to W^+W^-$ observables are related to $h\to Z\ell^+\ell^-$, when the precision electroweak constraints in the from of oblique parameters $\De_{1,2,3}\to0$ are imposed. The latter can be done consistently when we restrict ourselves to the 16-dimensional subspace of the SMEFT parameter space that characterizes universal theories. Our conclusion differs from that in~\cite{Trott:2014dma}, where a stronger restriction is placed on the SMEFT parameter space (the ``strong LEP bound limit'') that is however not required for the utility of the oblique parameters, and has the effect of decoupling the correlations shown here.

Of course, a separate issue is whether taking the limit $\De_{1,2,3}\to0$ as motivated by precision electroweak constraints is justified in TGC extractions. In the case of $e^+e^-\to W^+W^-$ at LEP2, which dominates the current anomalous TGC constraints, we find (also with the differential cross section $\frac{d\sigma}{d\cos\theta}$ taken into account) that the answer is positive, in the sense that in almost the entire phase space, the possible contributions from $\De_{1,2,3}$, as constrained by the oblique parameters analyses, are smaller than the contributions from the anomalous TGCs, when the latter saturate the upper bounds derived from $e^+e^-\to W^+W^-$ data assuming $\De_{1,2,3}\to0$. The same conclusion holds also for nonuniversal theories, if one assumes the invisible $Z$ decay width is equivalent to $\Gamma_{Z\to\bar\nu\nu}$ [so that $\dgL^{Wl}$ is strongly constrained from $\dgL^{Ze}$ and $\dgL^{Z\nu}$ by \eqref{couprel-ew}]. But in this case, one should use the precision electroweak constraints in the form of per-mil-level bounds on $\dbNP m_W$, $\dbNP\Gamma_{Z\to\ell^+\ell^-}$, $\dbNP\Gamma_{Z\to\bar\nu\nu}$, $\dbNP\sin^2\theta_{\text{eff}}$ instead of the oblique parameters. We remark, however, that the situation may change at future high-precision measurements of TGCs. A detailed analysis will be presented in a future publication.

\section{Conclusions}
\label{sec:conclusions}

While it is often desirable to simplify the indirect searches for BSM physics by introducing model-independent frameworks, it is important to understand the range of applicability of each framework so as not to use a framework to constrain BSM theories where it does not apply. As a historically influential example, oblique parameters analyses in general can only be used to connect precision electroweak data to universal theories, where it is possible to shuffle all the indirect BSM effects, or at least the dominant ones, into the bosonic sector. On the other hand, the SMEFT, as the modern approach to model-independently study BSM effects on precision observables, is completely general (assuming the absence of light new states). Caution is needed when connecting the two frameworks, to ensure the analysis is consistent and basis-independent. In particular, one should not naively write down the oblique parameters from the vector boson self-energy corrections in a specific basis for the most general SMEFT, or use the reported bounds on the oblique parameters to constrain the full parameter space of the SMEFT.

In this paper we have presented a detailed EFT analysis of universal theories. As we have shown, universal theories can be unambiguously defined in any SMEFT basis, in terms of restrictions on the Wilson coefficients. When these restrictions are satisfied, the oblique parameters can be written in terms of the SMEFT Wilson coefficients in a basis-independent way. To completely characterize the SM deviations in universal theories, however, requires extending the oblique parameters formalism to 16 ``universal parameters'' that we have defined; see \eqref{Lutc}. Table~\ref{tab:utc} shows how these universal parameters should be written down in each SMEFT basis. While the electroweak oblique parameters, especially $\hat S$ and $\hat T$, have been under intensive study historically due to the strong precision electroweak constraints, they do not have a special status in the complete characterization of universal theories. As we begin to push the precision frontier to the Higgs sector, more universal parameters have become (or will soon become) accessible, although with perhaps lower precisions at the present stage (or in the near future).

The universal pattern of SM deviations in universal theories becomes transparent when the analysis is connected to the Higgs basis framework, and the Higgs basis couplings are expressed in terms of the universal parameters as in table~\ref{tab:hb}. This demonstrates how the otherwise independent effective couplings are related in universal theories, as summarized in section~\ref{sec:hb-cor}. Further, we have illustrated two example applications to phenomenology -- corrections to the precision electroweak observables, and the connections between anomalous TGCs and Higgs couplings. All our analyses have been done at leading order in the new physics contributions. We will discuss RG effects in universal theories in a follow-up paper~\cite{followup}.

As precision analyses continue to guide us in the search for new physics, the importance of ensuring theory consistency will grow as more data, especially in the Higgs sector, become available. Our analysis constitutes an effort toward this aim.

\acknowledgments
We thank Michael Trott and Cen Zhang for useful discussions. This work is supported in part the the U.S.\ Department of Energy under grant DE-SC0007859.


\appendix
\section{Notation and useful formulas}
\label{sec:app}

Our notation is such that
\beqa
\L_\SM &=& -\frac{1}{4}G^A_{\mu\nu}G^{A\mu\nu} -\frac{1}{4}W^a_{\mu\nu}W^{a\mu\nu} -\frac{1}{4}B_{\mu\nu}B^{\mu\nu} +|D_\mu H|^2 +\lambda v^2|H|^2 -\lambda|H|^4 \CR
&& +\sum_{f\in\{q,l,u,d,e\}}i\bar f\gamma^\mu D_\mu f -\bigl[(\bar u y_u^\dagger q_\beta \epsilon^{\beta\alpha} + \bar q^\alpha \VCKM y_d d + \bar l^\alpha y_e e)H_\alpha+\text{h.c.}\bigr].
\eeqan
Denoting in general an antisymmetric tensor by $(\dots)_{[\mu,\nu]}\equiv(\dots)_{\mu\nu}-(\dots)_{\nu\mu}$, we have $G^A_{\mu\nu} = \partial_{[\mu,} G^A_{\nu]}+g_sf^{ABC}G^B_\mu G^C_\nu$, $W^a_{\mu\nu}= \partial_{[\mu,} W^a_{\nu]}+g\epsilon^{abc}W^b_\mu W^c_\nu$, $B_{\mu\nu}= \partial_{[\mu,} B_{\nu]}$. The $SU(2)_L$ doublets $q=(u_L,d_L)$, $l=(\nu,e_L)$, and the $SU(2)_L$ singlets $u=u_R$, $d=d_R$, $e=e_R$. All the gauge-eigenstate fermion fields are also mass eigenstates except $d_L=\VCKM d'_L$ where $d'_L$ is a mass eigenstate. For $f=q$, $D_\mu=\partial_\mu-ig_sT^AG^A_\mu-ig\frac{\sigma^a}{2}W^a_\mu-ig'Y_fB_\mu$ with $[T^A,T^B]=if^{ABC}T^C$, $[\frac{\sigma^a}{2},\frac{\sigma^b}{2}]=i\epsilon^{abc}\frac{\sigma^c}{2}$; the $SU(3)_c$ and/or $SU(2)_L$ pieces are absent for other fermion fields neutral under these gauge groups. In the last term, $\alpha$ and $\beta$ are $SU(2)_L$ indices of the doublet fields, while generation indices are implicitly summed over; the $3\times3$ Yukawa matrices in generation space $y_u$, $y_d$, $y_e$ are diagonal and real.

In the unitary gauge, $H=\frac{1}{\sqrt{2}}(0,v+h)$ where $h$ is the physical Higgs boson. Electroweak symmetry breaking mixes $W^3$ and $B$ to form the mass eigenstates
\beq
Z_\mu = \cw W^3_\mu -\sw B_\mu,\quad A_\mu = \sw W^3_\mu +\cw B_\mu,
\eeqn
where
\beq
\cw = \frac{g}{\sqrt{g^2+g'^2}} = \frac{e}{g'},\quad \sw = \frac{g'}{\sqrt{g^2+g'^2}}= \frac{e}{g}.
\eeqn
Inversely,
\beq
W^3_\mu = \cw Z_\mu +\sw A_\mu,\quad B_\mu = -\sw Z_\mu +\cw A_\mu.
\eeqn
The charged gauge bosons $W^\pm$, on the other hand, are related to $W^{1,2}$ by
\beq
W^\pm_\mu = \frac{1}{\sqrt{2}}(W^1_\mu \mp iW^2_\mu);\quad W^1_\mu = \frac{1}{\sqrt{2}}(W^+_\mu +W^-_\mu),\,\, W^2_\mu = \frac{i}{\sqrt{2}}(W^+_\mu -W^-_\mu).
\eeqn
The mass-eigenstate field strengths are defined by
\beq
W^\pm_{\mu\nu} = \partial_{[\mu,}W^\pm_{\nu]},\quad Z_{\mu\nu} = \partial_{[\mu,}Z_{\nu]},\quad A_{\mu\nu} = \partial_{[\mu,}A_{\nu]}.
\eeqn

The gauge interactions of the SM fermions read
\beqa
&& G^{A\mu}J_{G\mu}^A + W^{a\mu}J_{W\mu}^a + B^\mu J_{B\mu} \CR
&=& g_s G^A_\mu \sum_{f\in\{q,u,d\}}\bar f \gamma^\mu T^A f +\frac{g}{\sqrt{2}} \Bigl(W^+_\mu \sum_{f\in\{q,l\}} \bar f \gamma^\mu \sigma^+ f +\text{h.c.}\Bigr) \CR
&& +\sum_{f\in\{q,l,u,d,e\}} \Bigl[\frac{g}{\cw}Z_\mu(T^3_f-Q_f\sw^2)+eA_\mu Q_f\Bigr]\bar f \gamma^\mu f \CR
&=& g_s G^A_\mu \sum_{f\in\{u_L,d_L,u_R,d_R\}} \bar f \gamma^\mu T^A f +\frac{g}{\sqrt{2}} \bigl[W^+_\mu (\bar u_L \gamma^\mu \VCKM d'_L +\bar\nu\gamma^\mu e_L) +\text{h.c.}\bigr] \CR
&& +\sum_f \Bigl[\frac{g}{\cw}Z_\mu(T^3_f-Q_f\sw^2)+eA_\mu Q_f\Bigr]\bar f \gamma^\mu f,
\eeqa{LVffSM}
where $\sigma^+=(\sigma^1+i\sigma^2)/2$. The last sum is over $f\in\{u_L,u_R,d_L,d_R,e_L,e_R,\nu\}$, with $T^3_f=\{\frac{1}{2},0,-\frac{1}{2},0,-\frac{1}{2},0,\frac{1}{2}\}$, respectively. $Q_f=T^3_f+Y_f$ with $Y_f$ given in \eqref{Yf}.

It is useful to know the following Fierz rearrangement formulas,
\bseq
\beqa
(\bar f_{1L}\gamma_\mu f_{2L})(\bar f_{3L}\gamma_\mu f_{4L}) &=& (\bar f_{1L}\gamma_\mu f_{4L})(\bar f_{3L}\gamma_\mu f_{2L}), \\
(\bar f_{1L}f_{2R})(\bar f_{3R}f_{4L}) &=& -\frac{1}{2}(\bar f_{1L}\gamma_\mu f_{4L})(\bar f_{3R}\gamma_\mu f_{2R}).
\eeqan
\eseq{Fierz}
The same identities hold with $L\leftrightarrow R$. Note that the $f$'s in these equations are anticommuting fields dependent on the spacetime coordinate $x^\mu$; if these formulas are derived for the momentum-space spinors $u_{L,R}(p)$, $v_{L,R}(p)$, which are commuting, the right hand sides should be multiplied by $(-1)$. Also, the following group-theoretic identities are often used when reducing operators,
\bseq
\beqa
\sigma^a_{\alpha\beta}\sigma^a_{\gamma\delta} &=& 2\delta_{\alpha\delta}\delta_{\gamma\beta} -\delta_{\alpha\beta}\delta_{\gamma\delta}, \\
T^A_{ab}T^A_{cd} &=& \frac{1}{2}\delta_{ad}\delta_{cb} -\frac{1}{6}\delta_{ab}\delta_{cd}.
\eeqan
\eseq{groupid}



\end{document}